\documentclass[prb,longbibliography,twocolumn,aps,superscriptaddress,10pt]{revtex4-2}
\usepackage[utf8]{inputenc}
\usepackage{graphicx}
\usepackage{color}
\usepackage{float}
\usepackage[bookmarks=false,linkcolor=blue,urlcolor=blue,colorlinks,citecolor=blue]{hyperref}
\usepackage{amsmath,mathtools}
\usepackage{dsfont}
\usepackage{amssymb}
\usepackage{soul}
\newcommand{\one}{\mathds{1}}

\newcommand{\Tr}{\mathop{\text{Tr}}\nolimits}
\newcommand{\ket}[1]{\vert {#1}\rangle}
\newcommand{\bra}[1]{\langle{#1}|}
\newcommand{\Bth}{B_{\text{th}}}
\newcommand{\BQPC}{B_{\text{QPC}}}
\newcommand{\Eq}[1]{Eq.~(\ref{#1})}
\newcommand{\Eqs}[1]{Eqs.~(\ref{#1})}
\newcommand{\BrackEq}[1]{[Eq.~(\ref{#1})]}
\newcommand{\BrackSec}[1]{[Sec.~(\ref{#1})]}
\newcommand{\eq}[1]{(\ref{#1})}
\newcommand{\Fig}[1]{Fig.~\ref{#1}}

\newcommand{\Figs}[1]{Figs.~\ref{#1}}
\newcommand{\Sec}[1]{Sec.~\ref{#1}}
\newcommand{\Refe}[1]{Ref.~\onlinecite{#1}}
\newcommand{\Refs}[1]{Refs.~\onlinecite{#1}}
\newcommand{\nbrack}[1]{\left(#1\right)}
\newcommand{\sqbrack}[1]{\left[#1\right]}

\newcommand{\sqlbrack}[1]{\left[#1\right.}

\newcommand{\sqrbrack}[1]{\left.#1\right]}

\newcommand{\lrsepa}{\quad,\quad}
\newcommand{\sumsub}[2]{\sum_{\substack{{#1}\\{#2}}}}
\begin{document}
\title{Multi-level effects in quantum-dot based parity-to-charge conversion of Majorana box qubits}

\author{Jens Schulenborg}
\affiliation{
  Center for Quantum Devices, Niels Bohr Institute, University of Copenhagen, 2100 Copenhagen, Denmark
}
\author{Michele Burrello}
\affiliation{
  Center for Quantum Devices, Niels Bohr Institute, University of Copenhagen, 2100 Copenhagen, Denmark
}
\affiliation{
 Niels Bohr International Academy, Niels Bohr Institute, University of Copenhagen, 2100 Copenhagen, Denmark
}
\author{Martin Leijnse}
\affiliation{
    Solid State Physics and NanoLund, Lund University, Box 118, S-221 00 Lund, Sweden
}
\affiliation{
  Center for Quantum Devices, Niels Bohr Institute, University of Copenhagen, 2100 Copenhagen, Denmark
}
\author{Karsten Flensberg}
\affiliation{
  Center for Quantum Devices, Niels Bohr Institute, University of Copenhagen, 2100 Copenhagen, Denmark
}
             
\begin{abstract}
 Quantum-dot based parity-to-charge conversion is a promising method for reading out quantum information encoded nonlocally into pairs of Majorana zero modes. To obtain a sizable parity-to-charge visibility, it is crucial to tune the relative phase of the tunnel couplings between the dot and the Majorana modes appropriately. However, in the presence of multiple quasi-degenerate dot orbitals, it is in general not experimentally feasible to tune all couplings individually. This paper shows that such configurations could make it difficult to avoid a destructive multi-orbital interference effect that substantially reduces the read-out visibility. We analyze this effect using a Lindblad quantum master equation. This exposes how the experimentally relevant system parameters enhance or suppress the visibility when strong charging energy, measurement dissipation and, most importantly, multi-orbital interference is accounted for. In particular, we find that an intermediate-time readout could mitigate some of the interference-related visibility reductions affecting the stationary limit.
\end{abstract}
\maketitle

\section{Introduction}\label{sec1}

Topological superconductors hosting Majorana zero-energy modes~\cite{Kitaev2001,Lutchyn2010Aug,Oreg2010Oct,Mourik2012May,Das2012Nov,Rokhinson2012Sep,Wiedenmann2016Jan,Deng2016Dec,Zhang2018Mar} provide a potential platform to realize topologically protected states whose fermion parity can store quantum information non-locally~\cite{Kitaev2003Jan}. The non-local nature of Majorana modes makes them a good candidate for the development of qubits that are resilient against local noise and perturbations~\cite{Nayak2008Sep,Leijnse2012Nov,Beenakker2013Mar,Aguado2017Oct,Beenakker2020Aug}, and can be adopted as building blocks for quantum memories and quantum information processing architectures \cite{Terhal2012Jun,Hyart2013Jul,Sarma2015Oct,Aasen2016Aug,Landau2016Feb,Plugge2016Nov,Plugge2017Jan,Karzig2017Jun,Litinski2017Sep}.

A crucial element in any quantum information application based on Majorana modes is the ability to read out the fermionic parity of a pair of Majoranas. Most of the above cited proposals are based on hybrid semiconductor-superconductor platforms, where the Majorana modes are fixed to specific locations, such as, e.g., the ends of topological superconducting wires~\cite{Kitaev2001,Lutchyn2010Aug,Oreg2010Oct,Mourik2012May}.
Among many suggested readout methods~\cite{Ohm2015Feb,Gharavi2016Oct,Malciu2018Oct,Li2018Nov,Grimsmo2019Jun,Szechenyi2020Jun,Munk2020Aug,Steiner2020Aug,Smith2020Nov}, this allows for what we call \emph{parity-to-charge conversion}~\cite{Flensberg2011Mar,Aasen2016Aug,Gharavi2016Oct,Prada2017Aug,Clarke2017Nov,Karzig2017Jun,Szechenyi2020Jun,Munk2020Aug,Steiner2020Aug}: a readout protocol translating the non-local parity of a Majorana pair into a local charge or current, via a suitable coupling to the Majoranas.

\begin{figure}[t!!]
\includegraphics[width=\linewidth]{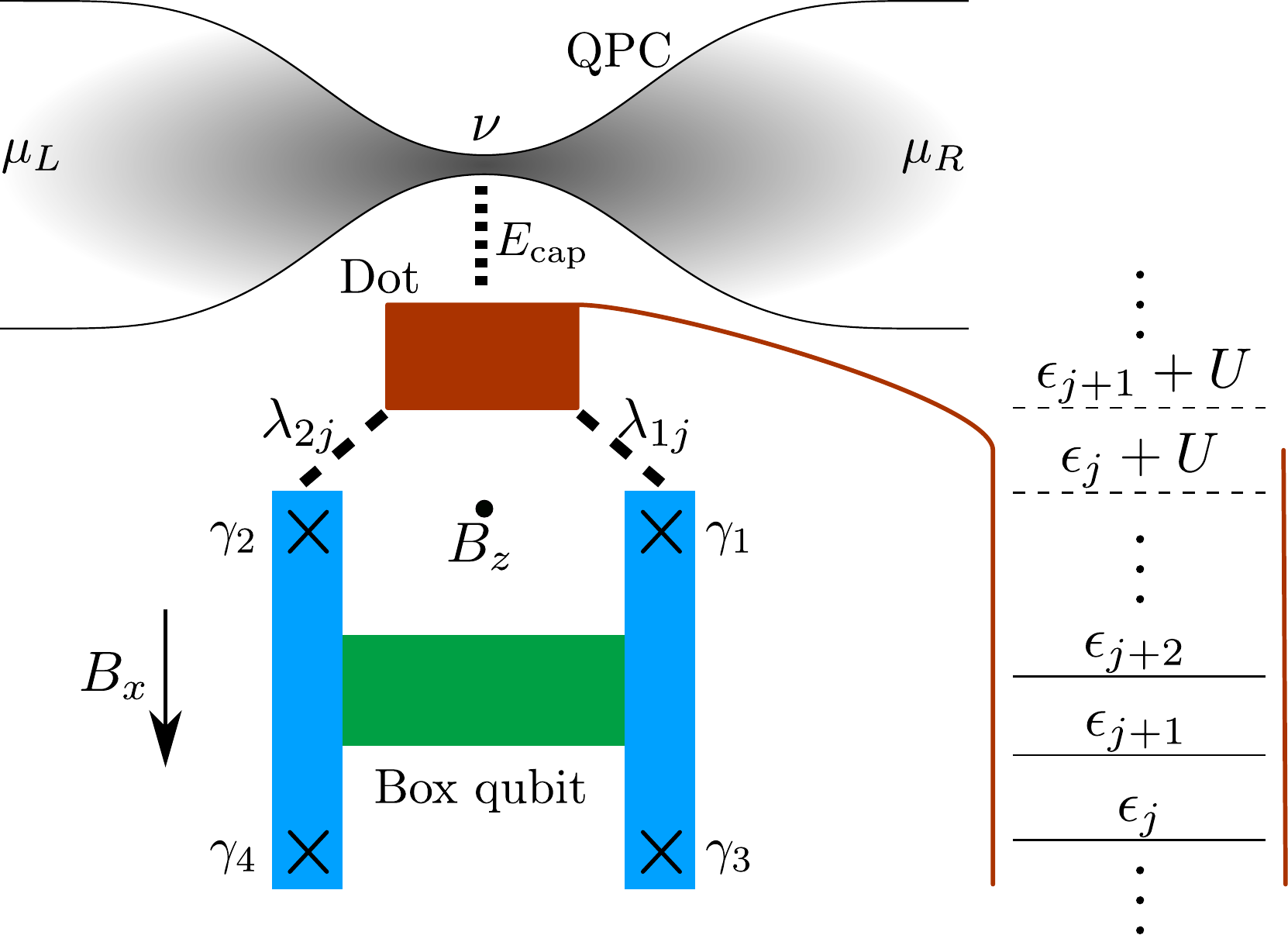}
\caption{The system of interest consists of a Majorana box qubit, a tunnel coupled multi-level dot, and a quantum point contact (QPC) charge detector capacitively coupled to the dot. The qubit is constituted by 4 Majoranas $\gamma_i$ at the ends of two topological superconducting wires (blue). The wires are bridged by an s-wave superconductor (green) and a parallel magnetic field with Zeeman energy $B_x$ is applied. The Majoranas $\gamma_{1,2}$ couple with strengths $\lambda_{1j},\lambda_{2j}$ to multiple single-particle states $j$ in a quantum dot (red) with energies $\epsilon_j$. The relative coupling phases $\text{arg}(\lambda_{2j}) - \text{arg}(\lambda_{1j})$ can, collectively, be shifted via a tunable magnetic flux $\sim B_z$ orthogonal to the dot-qubit plane. The dot and tetron are subject to strong local Coulomb interaction $U$ restricting the dynamics to a single charge hopping between Majorana qubit and dot. The QPC sensor is potential biased with $\Delta\mu = \mu_{\text{L}} - \mu_{\text{R}}$, and it has a density of states $\nu$ in the constriction close to the dot. Here, the QPC charge density weakly, capacitively couples to the dot charge with strength $E_{\text{cap}}$, meaning $g \sim (E_{\text{cap}}\nu)^2 \ll 1$.  \label{fig_setup}}
\end{figure}

A controlled way to perform parity-to-charge conversion is through a quantum dot: a tunnel coupling between dot and Majorana pair breaks the energetic fermion-parity degeneracy of the latter, and this in return leads to a parity-dependent ground-state dot occupation. The use of tunnel-coupled quantum dots is experimentally well-established, and presents many technological advantages. Quantum dots are experimentally feasible in hybrid topological superconductor devices~\cite{Deng2014Dec,Deng2016Dec,Albrecht2016Mar,Lutchyn2018May,Deng2018Aug,Razmadze2020Sep} and their properties can typically be accurately tuned through suitable voltage gates. Furthermore, many techniques to measure their occupation have been successfully applied, including readout via coupled electromagnetic resonators~\cite{Yoshie2004Nov,Reithmaier2004Nov,Delbecq2011Dec,Frey2012Jan,Petersson2012Oct,Xiang2013Apr,Stockklauser2017Mar,Burkard2020Mar,Deng2020Apr}, metallic islands and sensor dots~\cite{Schoelkopf1998May,Lu2003May,Fujisawa2004Mar,Bylander2005Mar,Buehler2005Apr,Barthel2010Apr,Maisi2011May}, or quantum point contacts (QPCs)~\cite{Field1993Mar,Elzerman2003Apr,Elzerman2004Jul,Ihn2009Sep,Barthel2009Oct}.

Such readout schemes, however, introduce additional challenges that must be tackled to obtain reliable measurements. First, the tunneling-induced energy splitting influences not only the ground state, but the entire dot-Majorana dynamics, and in particular measurement-related decay and decoherence. The result is, as shown below, that the actual parity-dependence of the dot occupation can deviate substantially from the ground-state expectation. Second, the induced energy splitting depends on the \emph{interference}, and hence the relative phase between the dot-Majorana tunnel couplings~\cite{Karzig2017Jun,Szechenyi2020Jun,Munk2020Aug,Steiner2020Aug}. As long as only a single dot level is involved, a tunable magnetic flux may provide a control knob to optimize this phase. However, quantum dots in practice often exhibit near-degenerate orbitals which could simultaneously couple to the Majoranas. In such situations, multiple dot levels may potentially interfere destructively with respect to the net induced energy splitting, leading to a significant visibility loss.

This paper theoretically assesses the effect of multiple quantum dot orbitals on the performance of a quantum-dot based parity-to-charge conversion scheme as sketched in \Fig{fig_setup}. Using a Lindblad-type quantum master equation~\cite{Gorini1976May,Lindblad1976Jun} based on the \emph{coherent approximation}~\cite{Kirsanskas2018Jan,Mozgunov2020Feb,Nathan2020Sep,Kleinherbers2020Mar}, we study the parity-dependence of the multi-orbital dot occupation in the presence of strong dot charging energy and measurement-induced dissipation.
More precisely, after establishing the theoretical model and method to describe quantum-dot based parity-to-charge conversion, \Sec{sec2} first reviews the single-level dot case detailed in \Refs{Munk2020Aug,Steiner2020Aug}. The main differences connected to multiple orbitals are examined by introducing a second dot level in \Sec{sec3a}, and finally by extending the analysis to many levels in \Sec{sec3c}. Section~\ref{sec3b} bridges the sections by providing typical estimates for dot-level splittings and relative tunnel coupling phases, given the example of two Majoranas overlapping with a two-dimensional quantum well.

\section{Parity-to-charge conversion}
\label{sec2}

\subsection{Model and readout principle}\label{sec2a}

The system of interest is sketched in \Fig{fig_setup}: the Majorana box qubit, also known as tetron,~\cite{Plugge2017Jan,Karzig2017Jun}, is formed by two topological superconducting wires, each hosting a Majorana mode $\gamma$ at both ends. A topologically trivial s-wave superconductor bridges the two wires. The common superconducting gap $\Delta$ is assumed to be large enough for the Majorana excitations to be treated independently from the quasi-particle continuum.
The qubit states $\ket{n_{12}n_{34}}$ are encoded into the occupations $n_{12} = f^\dagger_{12}f_{12}$ and $n_{34} = f^\dagger_{34}f_{34}$ of the fermions $f^\dagger_{12} = (\gamma_1 - i\gamma_2)/\sqrt{2}$ and $f^\dagger_{34} = (\gamma_3 - i\gamma_4)/\sqrt{2}$, where we adopt the normalization convention $\left\{\gamma_i,\gamma_j\right\} = \delta_{ij}$. We define two logical states as $\ket{0} = \ket{n_{12} = 0, n_{34} = 0}$ with $f_{12}\ket{0} = f_{34}\ket{0} = 0$, and $\ket{1} = \ket{n_{12}=1,n_{34}=1} = f^\dagger_{12}f^\dagger_{34}\ket{0}$. Both are chosen to lie in the even fermionic parity sector, assuming that the time between parity flips due to, e.g., quasi-particle poisoning, is long enough to reliably prepare such states. 

The Majoranas $\gamma_1,\gamma_2$ are tunnel coupled to a quantum dot, yielding the Hamiltonian
\begin{align}
 H &= \sum_j^M\epsilon_j n_j +\sum_j^M\sum_{i=1,2}\gamma_i\sqbrack{\lambda_{ij} d_j - \lambda_{ij}^* d^\dagger_j}\label{eq_hamiltonian}
\end{align}
The key difference to previous works~\cite{Flensberg2011Mar,Karzig2017Jun,Khindanov2020Jul,Munk2020Aug,Steiner2020Aug} is that here, we do not assume the dot to be described by a single electronic level. We instead consider a more realistic multi-orbital scenario with $M$ dot states $j$ entering the dynamics. Each level couples to Majorana $i = 1,2$ with mutually different coupling constants $\lambda_{ij}$. A large charging energy in both the quantum dot and the floating tetron limits the parity exchange to a single charge moving to or from the dot. Thus, we consider the Hamiltonian \eq{eq_hamiltonian} only in the subspace of excess dot occupation $N = \sum_j^M n_j \leq 1$, where the $n_j = d^\dagger_j d_j$ are the individual occupations created and annihilated by $d^\dagger_j$ and $d_j$. Any further level that would lie below $\epsilon_1$ and above $\epsilon_M$ is assumed to be constantly occupied or empty, and is hence not included in \Eq{eq_hamiltonian}. The remaining $2(M+1)$ total even-parity many-body basis states in the dynamics are $\ket{N = 0; n_{12} = n_{34} = 0}$ and $\{\ket{N = 1; n_j = 1; n_{12} = 1, n_{34} = 0}\}_j$ with even subparity $s = (-\one)^{N + n_{12}} = 1$, and $\ket{N = 0; n_{12} = n_{34} = 1}$ as well as $\{\ket{N = 1; n_j = 1; n_{12} = 0, n_{34} = 1}\}_j$ with odd subparity $s = (-\one)^{N + n_{12}} = -1$. 

The tunnel coupling energetically splits the two otherwise degenerate Majorana states corresponding to $n_{12} = 0$ and $n_{12} = 1$, and thereby enables the state readout. This readout crucially relies on the fact that the tunneling conserves the \emph{subparity} $s = (-\one)^{N + n_{12}}$. Initially prepared qubit states $\ket{0} = \ket{n_{12} = n_{34} = 0}$ and, respectively, $\ket{1} = \ket{n_{12} = n_{34} = 1}$, therefore evolve in mutually disconnected subparity subspaces as long as the \emph{total} fermionic parity is constant: given $(-\one)^{N + n_{12} + n_{34}} = 1$, the initial state $\ket{0}$ stays in the \emph{even} sector $s = 1$, and $\ket{1}$ in the odd sector $s = -1$. Hence, after calibration, measuring an $s$-dependent dot observable allows to infer the qubit state prepared prior to the onset of the dot-Majorana coupling, see details in \Refs{Munk2020Aug,Steiner2020Aug}.

The central point in this work is that the $s$-dependence of the dot occupation is affected by how the tunnel couplings $\lambda_{ij} = |\lambda_{ij}|e^{i\phi_{ij}}$ differ between different orbitals $j'\neq j$. For only a single orbital $j$, external electric fields would in principle allow to tune the couplings close to an optimum of symmetric amplitudes~\cite{Munk2020Aug}, $|\lambda_{2j}| = |\lambda_{1j}|$. An external magnetic flux $\sim B_z$ perpendicular to the dot-tetron plane can likewise be used to optimize the phase difference $\phi_{2j} - \phi_{1j}$. However, if more dot orbitals $j'\neq j$ become relevant, the applied fields only act collectively on the couplings. This implies a potentially unavoidable orbital dependence in the overall amplitudes, $|\lambda_{ij'}|$, in the amplitude asymmetries with respect to the Majoranas, $||\lambda_{2j'}| - |\lambda_{1j'}||$, and, most importantly, in the relative phases $\phi_{2j'} - \phi_{1j'}$.
Just as a reduction of the overall coupling amplitude, also Majorana-dependent amplitude asymmetries reduce the interaction strength between dot orbitals and the Majoranas, see \Refe{Munk2020Aug}. This agrees with the well-known intuition that Majoranas can only interact in pairs with the outside world. In particular, it means that if one, e.g., increases the amplitude asymmetry of orbital $j'$ compared to orbital $j$ to $||\lambda_{2j'}| - |\lambda_{1j'}|| \gg ||\lambda_{2j}| - |\lambda_{1j}||$, the influence of $j'$ on the Majoranas reduces significantly compared to $j$. 
 
 Our study, however, focuses on the less straightforward situation in which multiple dot orbitals couple similarly in amplitude to both Majoranas, but \emph{asymmetrically in phase}:
 \begin{equation}
 \phi_{2j} - \phi_{1j} \neq \phi_{2j'} - \phi_{1j'} \text{ for } j'\neq j.\label{eq_phase_asymmetry}
\end{equation}
In such situations, the dot orbitals can interfere coherently in their interaction with the Majoranas depending on the phase asymmetry \eq{eq_phase_asymmetry}. Indeed, our key result shown below is that this orbital-dependent interference can substantially reduce the qubit-state visibility. To fully understand this, we, however, first review parity-to-charge conversion in the ideal, single-level case.

\subsection{Ground-state readout in single-level system}\label{sec2b}

\begin{figure}[t!!]
\includegraphics[width=\linewidth]{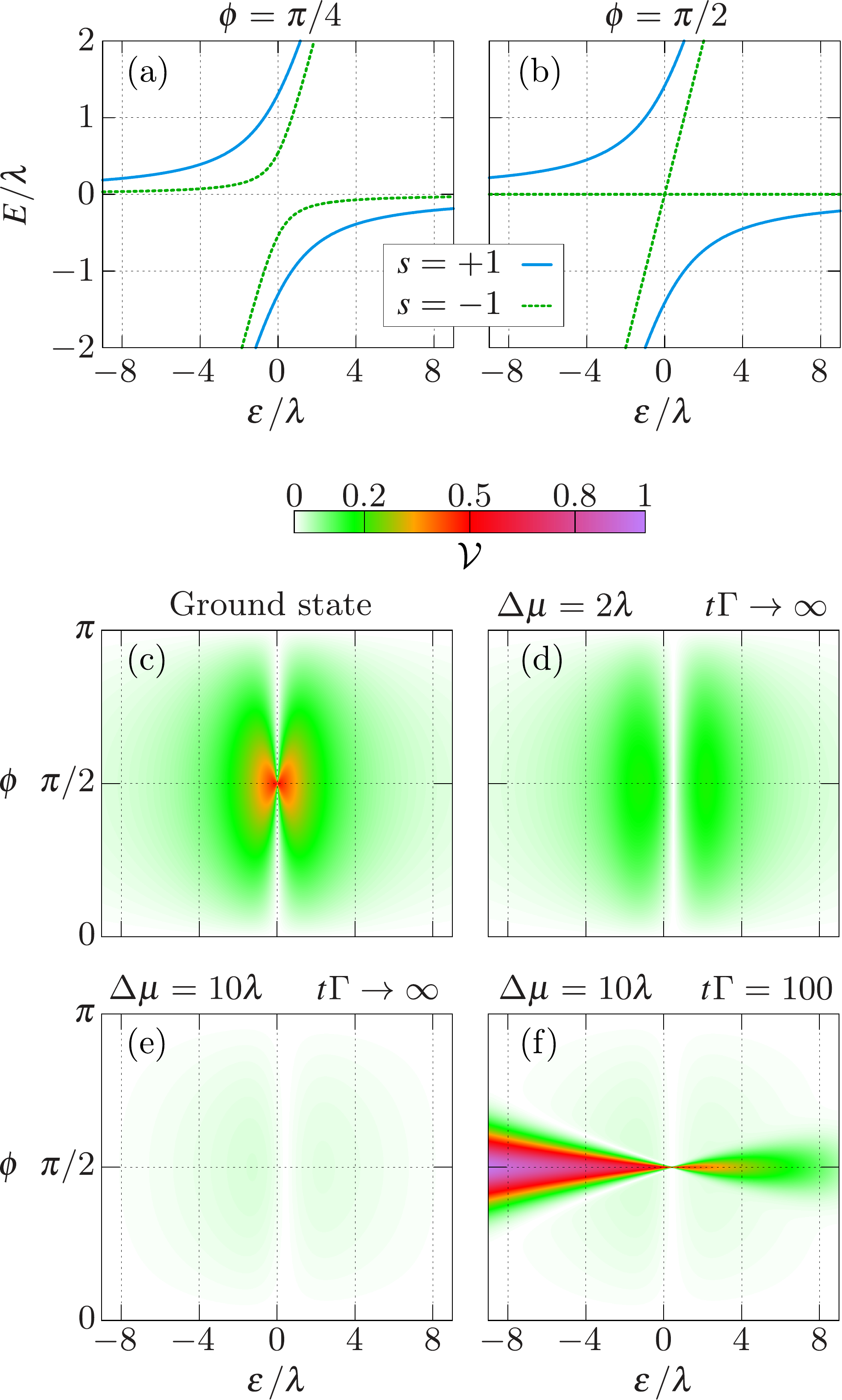}
\caption{(a,b) $\epsilon$-dependent two-body spectra of the Hamiltonian \eq{eq_hamiltonian_single} for $\phi = \pi/4$ (a) and $\phi = \pi/2$ (b). (c-f) Charge visibility of a single dot level as a function of level detuning $\epsilon$ with respect to the particle-hole symmetry point $\epsilon = 0$ and tunnel coupling phase $\phi$ \BrackEq{eq_coupling_phase}. (c) Visibility \eq{eq_visbility} in the $s$-dependent ground states of Hamiltonian \eq{eq_hamiltonian_single}. (d-f) Visibility \eq{eq_visbility_time} with coupled QPC sensor for different potential bias $\Delta\mu$ in the stationary limit and for a large finite time, where $g = 0.05$, $\Gamma = g\lambda$, $T = \lambda/5$ and $\omega_c = 20\lambda$ \BrackEq{eq_qpc_bath_correlation_fourier_final}. In (f), the dot is initially empty, $N(t = 0) = 0$. Note that phases $\pi < \phi \lesssim 2\pi$ are equivalent to phases $0 < \phi \lesssim \pi$ with $s \rightarrow -s$.   \label{fig_visibility_single}}
\end{figure}

As a simple example, we consider the occupation number $n = d^\dagger d$ of a single, effectively spinless dot level with detuning $\epsilon$ from zero energy,
and with tunnel couplings 
\begin{equation}
\lambda_1 = \lambda > 0 \text{ and } \lambda_2 = \lambda e^{i\phi}\label{eq_coupling_phase}.
\end{equation}
determined by an equal strength $\lambda$, but a difference in phase $\phi$ between Majorana 1 and 2.
The general Hamiltonian \eq{eq_hamiltonian} in this case simplifies to
\begin{align}
 H &\rightarrow \epsilon n + \sqrt{2}\lambda\sqlbrack{\cos\nbrack{\frac{\phi}{2} + \frac{\pi}{4}}\nbrack{f^\dagger_{12}\tilde{d} - f_{12}\tilde{d}^\dagger}}\label{eq_hamiltonian_single}\\
 &\phantom{\rightarrow \epsilon n + \sqrt{2}\lambda\sqlbrack{}}\sqrbrack{- i\cos\nbrack{\frac{\phi}{2} - \frac{\pi}{4}}\nbrack{f^\dagger_{12}\tilde{d}^\dagger + f_{12}\tilde{d}}}\notag
\end{align}
with $\tilde{d} = e^{i(\phi + \pi/2)/2}d$; the $\epsilon$-dependence of the corresponding many-body spectra are plotted in \Figs{fig_visibility_single}(a,b). The $\phi$-dependence of the even-subparity $(s = (-\mathds{1})^{N + n_{12}} = 1)$ pair creation and annihilation terms in the second line of \Eq{eq_hamiltonian_single} is shifted by $\pi/2$ with respect to the odd subparity $(s = -1)$ regular tunneling term in the first line. This implies a complementary $s$-dependence: for phases $\phi$ with a strong dot-Majorana coupling in the $s = +1$ sector, the coupling in the $s = -1$ sector is suppressed and vice versa. The energy splitting 
\begin{equation}
  E_{s=\pm} = \sqrt{\epsilon^2 +4\lambda^2 (1 + s\sin\phi)}\label{eq_splitting_single}
\end{equation}
between the many-body ground and excited state of the Hamiltonian \eq{eq_hamiltonian_single} thus depends on subparity $s$, as clearly visible in \Figs{fig_visibility_single}(a,b). This $s$-dependence translates to an $s$-dependent ground-state occupation $n_{\text{GS},s} = \bra{E_{\text{GS},s}}n\ket{E_{\text{GS},s}}$, and hence to a finite \emph{charge visibility}
\begin{equation}
\mathcal{V} = |\langle n\rangle_{s=+} - \langle n\rangle_{s=-}|\label{eq_visbility}
\end{equation}
in this ground state. The latter is plotted as a function of detuning $\epsilon$ and coupling phase $\phi$ in \Fig{fig_visibility_single}(c). As intuitively expected, phases $\phi \approx \pi/2$ and detunings $\epsilon \approx 0$ that maximize the relative $s$-dependence of the splitting \eq{eq_splitting_single} also yield maximal visibility (right at $\epsilon=0$ and $\phi=\pi/2$, there is no well-defined ground-state visibility due to the degeneracy for $s = -1$, see \Fig{fig_visibility_single}(f) and discussion below).

Importantly, however, the dot-Majorana subsystem does not generally assume the ground state when coupled to the measurement device. References \onlinecite{Munk2020Aug,Steiner2020Aug} show that when accounting for a capacitively coupled charge sensor -- such as a sensor quantum dot or quantum point contact as sketched in \Fig{fig_setup} -- the sensor becomes a dissipative environment for the dot and causes decay to a stationary state generally different from the ground state. Moreover, the initial state may be metastable with respect to the measurement induced decay, and its theoretical lifetime could exceed the experimental time limit set by other noise sources, such as quasi-particle poisoning. In this case, the ensemble averaged charge visibility can be better estimated from states at finite time prior to stationarity. To address these effects of time-dependent decay, we calculate the visibility with a Markovian quantum master equation for the reduced dot-qubit dynamics coupled to a sensor.

\subsection{Visibility in the presence of charge sensor}\label{sec2e}

For concreteness, we choose a quantum point contact (QPC) as the charge sensor that is weakly, capacitively coupled to the dot, as sketched in \Fig{fig_setup}. The corresponding total Hamiltonian reads
\begin{align}
H_{\text{tot}} &= H + H_{\text{env}} + H_{\text{coup}}\quad,\quad H_{\text{env}} = \sum_{\ell}\omega_\ell c^\dagger_\ell c_\ell\notag\\
H_{\text{coup}} &= E_{\text{cap}} N n_{\text{QPC}} \quad,\quad n_{\text{QPC}} = \sum_{\ell,\ell'}\xi_{\ell,\ell'}c^\dagger_{\ell}c_{\ell'}.
\end{align}
The QPC is represented as an effectively noninteracting environment $H_{\text{env}}$ with single-particle eigenstates $\ell$ of energy $\omega_\ell$ and corresponding creation and annihilation operators $c^\dagger_\ell,c_\ell$. The multi-index $\ell$ refers to wavenumber, spin, an index for the two terminals contacting at the QPC, and all further discrete degrees of freedom. The term $H_{\text{coup}}$ approximates the capacitive coupling between the charge density $n_{\text{QPC}}$ in the sensor region close to the dot, and the excess dot charge $N$. The internal wave function overlaps $\xi_{\ell,\ell'}$ between the two sides of the QPC are scaled by the overall capacitive coupling strength $E_{\text{cap}}$. We assume that within the relevant energy window, the product of the squared overlaps $|\xi_{\ell,\ell}|^2$ and density of states $\nu$ in the QPC constriction is approximately energy-independent. As argued in \Refe{Munk2020Aug}, we may then simplify $\xi_{\ell,\ell'} \rightarrow \xi$ and quantify the QPC-dot coupling strength by the single parameter $g = (\xi E_{\text{cap}}\nu)^2$. In this work, we are interested in the weak coupling regime, meaning $g \ll 1$. This translates to either a capacitive coupling $E_{\text{cap}}$ small compared to the inverse density of states $1/\nu$, or a sufficiently low transparency $\xi$.

In the presence of the QPC sensor, the dot and the two coupled Majoranas become an open quantum system whose dynamics are described by the reduced density operator $\rho(t) = \text{Tr}_{\text{env}}\sqbrack{\rho_{\text{tot}}(t)}$. Assuming internal relaxation in the contacts to be much faster than the typical QPC-dot interaction frequency $\sim g$, we obtain $\rho(t)$, ensemble-averaged over the environment states, from the Markovian quantum master equation~\cite{Bloch1946Oct,Redfield1965Jan,Gorini1976May,Lindblad1976Jun,Breuer2002,Gardiner2004}
\begin{equation}
 \partial_t\rho(t) = -i[H + \Lambda,\rho(t)] + L\rho(t) L^\dagger - \frac{1}{2}\{L^\dagger L,\rho(t)\}.\label{eq_master}
\end{equation}
The commutator $[H + \Lambda,\rho(t)]$ generates the effective local coherent dynamics of the dot and the two Majoranas, including the Lamb shift $\Lambda$ due to the coupled QPC. This Lamb shift is detailed in appendix~\ref{app_lamb}. There, we provide analytical expressions for $\Lambda$, and demonstrate that while mainly causing an irrelevant collective dot level shift for $t\rightarrow\infty$, the subparity-dependent Lamb shift does become quantitatively relevant at intermediate times. The most important effect of the QPC sensor on the subparity-dependent dynamics is, however, captured by the Lindbladian dissipator in the last two summands of \Eq{eq_master}.
The corresponding jump operator
\begin{equation}
 L = \sum_{\eta,\eta'}\sqrt{\BQPC(E_{\eta} - E_{\eta'})}\bra{\eta'} N\ket{\eta}\times \ket{\eta'}\bra{\eta}.\label{eq_jump_operator}
\end{equation}
represents environment-induced transitions between the different many-body eigenstates $\ket{\eta},\ket{\eta'}$ of the local Hamiltonian $H$ weighted by their energies $E_\eta,E_{\eta'}$. Compared to the commonly used secular approximation~\cite{Breuer2002,Gardiner2004}, the \emph{coherent approximation}~\cite{Kirsanskas2018Jan,Mozgunov2020Feb,Nathan2020Sep,Kleinherbers2020Mar} used in \Eq{eq_jump_operator} better describes the dissipative effect on local coherences, i.e. the off-diagonal elements of $\rho(t)$ in the eigenbasis of $H$. This is crucial, since such coherences become increasingly important in the here relevant \emph{multi-level} systems with small but finite energy splittings. The key point of  \Eq{eq_jump_operator} offering this benefit is that the jump operator itself -- and not its associated rate which here is normalized to one -- includes the energy weights via the QPC correlation function $\BQPC(\omega) = \int_{-\infty}^\infty dt B(t)e^{i\omega t}$. This function derives from the density-density correlator $B(t) \sim \langle (n_{\text{QPC}}(t) - \langle n_{\text{QPC}}(0)\rangle)(n_{\text{QPC}}(0) - \langle n_{\text{QPC}}(0)\rangle)\rangle$ in the interaction picture, $n_{\text{QPC}}(t) = e^{iH_{\text{env}}t}n_{\text{QPC}}e^{-iH_{\text{env}}t}$. For an initial grand-canonical environment with different chemical potentials for the two contacts, one obtains~\cite{Munk2020Aug}
 \begin{align}
\BQPC(\omega) &= \frac{2\Bth(\omega) + \Bth(\omega + \Delta\mu) + \Bth(\omega - \Delta\mu)}{2}\label{eq_qpc_bath_correlation_fourier_final}\notag\\
\Bth(\omega) &= g\frac{\pi }{2}\omega e^{-\frac{|\omega|}{\omega_c}}\left[\coth\nbrack{\frac{\omega}{2k_BT}} + 1\right],
\end{align}
where $T$ is the common temperature and $\Delta\mu$ is the potential difference between the two terminals at the QPC.
In other words, the sensor approximately behaves as two bosonic, bilinearly coupled Ohmic thermal baths with exponential energy cutoff. The cutoff frequency $\omega_c$ is determined by the QPC bandwidth.

Given a single dot level as in the previous section~\ref{sec2b}, \Fig{fig_visibility_single}(d-f) exemplify how the QPC affects the visibility, i.e., the difference 
\begin{equation}
 \mathcal{V}(t) = |\Tr\sqbrack{N\rho_+(t)} - \Tr\sqbrack{N\rho_-(t)}|\label{eq_visbility_time}
\end{equation}
in dot occupation of the two subparity sectors when evolving in time according to the master equation \eq{eq_master}
from an even subparity $(s = +)$ vs. an odd subparity $(s=-)$ initial state, $\rho(0)_\pm \rightarrow \rho_\pm(t)$. As a function of detuning $\epsilon/\lambda$ and coupling phase difference $\phi$, we plot both the stationary limit $t\rightarrow\infty$ $[$\Fig{fig_visibility_single}(d,e)$]$ and the situation at a finite time $t$ $[$\Fig{fig_visibility_single}(f)$]$. The stationary visibility at low to moderate potential bias $\Delta\mu \sim \lambda$ strongly resembles the ground-state visibility for most phases and detunings. The enhancement for phases $\phi \approx \pi/2$ and detunings $\epsilon/\lambda \approx 0$ in the ground state is, however, spoiled as a result of finite environment temperature and potential bias. Furthermore, the Lamb shift $\Lambda$ causes the zero-visibility line to slightly deviate from the particle-hole symmetric point $\epsilon = 0$ of the Hamiltonian $H$. At larger bias $\Delta\mu \gg \lambda$, the visibility mostly disappears in almost the entire parameter space. The reason for this becomes clear when analytically solving the master equation $-i[H + \Lambda,\rho_{\infty,s}] + L\rho_{\infty,s} L^\dagger - \frac{1}{2}\{L^\dagger L,\rho_{\infty,s}\} = 0$  for the stationary state $\rho_{\infty,s}$. With the effect of the Lamb shift $\Lambda$ mostly $s$-independent, and hence negligible for $t\rightarrow\infty$, the resulting stationary dot occupation $n_{\infty,s} = \Tr\sqbrack{N\rho_{\infty,s}}$ simplifies to
\begin{equation}
	n_{\infty,s} = \frac{1}{2} \left(1 - \frac{\epsilon}{E_s}\frac{\BQPC(E_s) - \BQPC(-E_s)}{\BQPC(E_s) + \BQPC(-E_s)}\right).\label{eq_stationary_charge}
\end{equation}
For biases considerably larger than the energy splitting \eq{eq_splitting_single}, $|\Delta\mu| > |E_s| \sim 3\lambda$ at moderate detuning $\epsilon \sim \lambda$, i.e. with both system transition energies within the bias window, we obtain $\BQPC(E_s) - \BQPC(-E_s) \rightarrow 0$ and thus $n_{\infty,s} \rightarrow 1/2$ \emph{independently of} subparity $s$. By providing enough energy for the system to constantly switch between ground and excited state \emph{regardless of subparity}, the QPC effectively prevents itself from discriminating $s$. In a concrete experiment with $\lambda \sim \text{GHz}$, this would correspond to potential differences $\Delta\mu \gg 10\,\mu\text{eV}$.

For the finite-time visibility in \Fig{fig_visibility_single}(f), we assume an empty dot at the beginning of the readout, $N(t = 0) = 0$, and choose $t\Gamma = tg\lambda = 100$ much larger than the inverse of the typical QPC coupling strength $\Gamma = g\lambda$. We hence expect near stationary visibilities for most system parameters. 
Nevertheless, the chosen time may still be small compared to typical quasi-particle poisoning times $t_{\text{QP}}$ in Majorana systems~\cite{Rainis2012May,Knapp2018Mar,Knapp2018Sep,Karzig2021Feb}, assuming, e.g., $\lambda \sim \text{GHz}$, $g = 0.05$ and $t_{\text{QP}} \sim 10-100\,\mu\text{s}$. 

According to \Fig{fig_visibility_single}(f), $\mathcal{V}$ at $t\Gamma = 100$ has indeed converged to its stationary value shown in \Fig{fig_visibility_single}(e) for most parameters $(\epsilon,\phi)$. There are, however, wedges along the $\epsilon$-axis around $\phi = \pi/2$ with strong deviations from stationarity. The tunneling phases $\phi\approx\phi/2$ not only maximize the energy splitting \eq{eq_splitting_single}, but also effectively decouple Majoranas and dot for $s = -1$, see \Fig{fig_visibility_single}(b):
\begin{equation}
 H \overset{\phi = \pi/2}{\rightarrow} \epsilon n + \sqrt{2}\lambda(f_{12} d - f^{\dagger}_{12} d^\dagger).\label{eq_subparity_decoupling}
\end{equation}
Phases $\phi \approx 3\pi/2$ would likewise decouple the system for $s = +1$. In either case, the initially prepared state becomes metastable in the corresponding subparity sector, and it provides a much larger visibility at intermediate times than in the stationary limit. Note that this not only holds for an initially empty dot, $N(t = 0) = 0$, but also for an initially filled dot, $N(t = 0) = 1$. The only difference is that while $N(0) = 0$ enhances visibilities primarily for negative detunings $\epsilon < 0$ favoring a filled dot in the stationary limit, $N(0) = 1$ would result in larger-than-stationary visibilities mostly for $\epsilon > 0$ eventually favoring an empty dot. The $\epsilon$-asymmetry observed in \Fig{fig_visibility_single}(f) would accordingly be inverted.

Crucially, the above described metastability is a property of the local Hamiltonian $H$, and hence does not critically depend on the QPC sensor. \emph{The suggested large-yet-finite time measurements could therefore optimize parity-to-charge readout}. This holds in particular if the bias $\Delta\mu$ is too large for sizable stationary visibility, as in \Fig{fig_visibility_single}(f), or when noise sources such as quasi-particle poisoning limit the measurement time.
We emphasize, however, that this interference effect requires a challenging fine tuning of the couplings towards symmetric amplitudes $|\lambda_1| \approx |\lambda_2|$ and phase $\phi \approx \pi/2$ with low parameter noise~\cite{Khindanov2020Jul}, as well as a good control of the initial dot state.
The latter also involves gate operations faster than the typical decay rate $\Gamma = g\lambda$, as $\epsilon$ needs to be varied from some large initial detuning $|\epsilon| \gg \lambda$ that strongly favors an empty or fully occupied dot level. Finally, the next section shows that time-dependent readout is also more susceptible to multi-orbital effects.  

\section{Multi-orbital effects}\label{sec3}

Let us now turn to the main results of this paper: the influence of multiple dot orbitals and, in particular, their interference. Since most of the qualitatively new effects already emerge by adding a second orbital to the single-level system in \Sec{sec2}, we first focus on such a two-level dot. The analysis is subsequently extended to more states.

\subsection{Two-level dot}
\label{sec3a}
In the case of $M = 2$ dot states $j=1,2$ with levels $\epsilon_1 = \epsilon - \Delta\epsilon/2, \epsilon_2 = \epsilon + \Delta\epsilon/2$, the subsystem Hamiltonian $H$ given in \Eq{eq_hamiltonian} contains altogether $4$ tunnel couplings $\lambda_{ij}$. As pointed out at the end of \Sec{sec2a}, we focus on multi-level interference effects, and hence on couplings with equal amplitudes $|\lambda_{ij}| = \lambda$ but possibly different phases:
\begin{gather}
 \lambda_{11} = \lambda_{12} = \lambda \,\,,\,\, \lambda_{21} = \lambda e^{i\phi} \,\,,\,\, \lambda_{22} = \lambda e^{i(\phi + \Delta\phi)}.\label{eq_tunnel_couplings}
\end{gather}
The dot-state dependence $\Delta\phi$ may have several physical origins, including state-dependent spatial wavelengths and spin-orbit coupling, as detailed in \Sec{sec3c}. Here, we simply take $\Delta\phi$ as an additional parameter.

\begin{figure}[t!!]
\includegraphics[width=\linewidth]{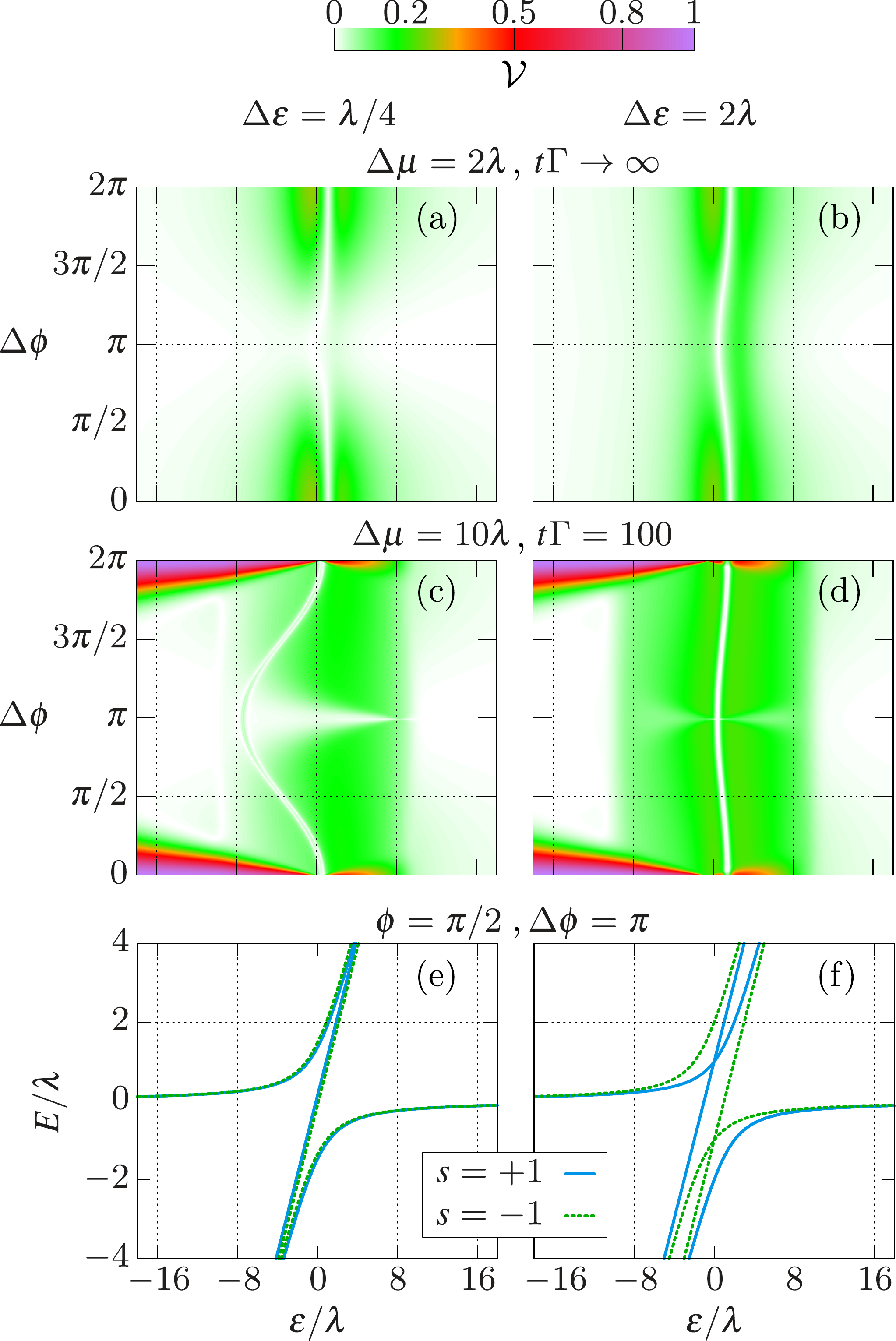}
\caption{(a-d) Charge visibility \eq{eq_visbility_time} with $M = 2$ dot levels as a function of the orbital dependence $\Delta\phi$ in the tunneling phase \BrackEq{eq_tunnel_couplings}, and of the average detuning $\epsilon = (\epsilon_1 + \epsilon_2)/2$. (e,f) $\epsilon$-dependence of the three lowest many-body energies of $H$ with $M=2$ \BrackEq{eq_hamiltonian}. In (c,d), we assume an initially empty dot, $N(t = 0) = 0$. Parameters common to panels (a-d) are $\phi = \pi/2$, $g = 0.05$, $T = \lambda/5$, $\omega_c = 20\lambda$. \label{fig_visibility_double}}
\end{figure}

In \Figs{fig_visibility_double}(a-d), the visibility \eq{eq_visbility_time} of the two-level dot in the presence of the QPC is shown as a function of $\epsilon$ and $\Delta\phi$ for the various stated parameter regimes, both in the stationary limit $[$\Fig{fig_visibility_double}(a,b)$]$ and at finite yet large time $[$\Fig{fig_visibility_double}(c,d)$]$. The phase $\phi$ is in all cases set to the above identified \BrackSec{sec2b} ideal visibility point $\phi = \pi/2$. 

First, we find a phase-dependent asymmetry with respect to the point $\epsilon = 0$ of zero average detuning, $\epsilon_1 = -\epsilon_2$. For both a single dot level or two equally coupled levels without charging energy, $\epsilon = 0$ corresponds to the particle-hole symmetry point at which the ground-state visibility disappears, $n_{s=+} = n_{s=-}$. However, next to the Lamb shift $\Lambda$, both an orbital-dependent coupling phase $\Delta\phi \neq 0$ and the here assumed large charging energy for two or more levels break this particle-hole symmetry. This not only moves the $\mathcal{V} = 0$ crossing away from $\epsilon = 0$; depending on $\Delta\phi$, and in clear contrast to the single-level case, it may also yield better visibility for detunings $\epsilon$ on one side of the crossing versus the other. 

As our main observation from \Figs{fig_visibility_double} in the stationary limit $t\rightarrow\infty$, we note a substantially reduced visibility for phase differences in a sizable interval around $\Delta\phi = \pi$ at small splitting $\Delta\epsilon = \lambda/4$. This is true when compared both to the case $\Delta\phi \approx 0$ and to the single-level-dot results in \Fig{fig_visibility_single}. The behavior is explained by the fact that for $\phi = \pi/2$ and $\Delta\phi \approx \pi$, all the many-body energies relevant for the dynamics are nearly independent of subparity $s$, see \Fig{fig_visibility_double}(e). For the single-level dot case treated in \Sec{sec2e}, the dot-Majorana level repulsion is the weaker for $s = +1$ the stronger it is for $s = -1$ \BrackEq{eq_hamiltonian_single}. By contrast, for two levels, a phase regime around $\Delta\phi = \pi$ opens up in which one dot level couples to the Majoranas mostly in the even $s$ sector, and the other dot level mostly in the odd $s$ sector; together, this cancels the effect on the net even-odd energy splitting. More loosely speaking, $\Delta\phi = \pi$ reverses the role of even and odd subparity for one of the two dot levels only, and thereby prevents the system from distinguishing between $s = +1$ and $s = -1$.
For a larger splitting $\Delta\epsilon = 2\lambda$, this $\Delta\phi$-dependence becomes much weaker $[$\Fig{fig_visibility_double}(b)$]$. The dot-Majorana coupling differs too much between the levels in this case, and the even-odd quasi-degeneracy in the spectrum disappears, see \Fig{fig_visibility_double}(f). An analogous effect is expected if the tunneling amplitudes $|\lambda_{ij}|$ would differ sizably between the two dot levels $j = 1,2$.

At intermediate time $t\Gamma = 100$ $[$\Fig{fig_visibility_double}(c,d)$]$ with an initially empty dot, $N(t = 0) = 0$, the strong metastable visibility enhancement of a single level dot around $\phi = \pi/2$   $[$\Fig{fig_visibility_single}(f)$]$ mostly disappears for $\Delta\phi$ deviating  from $0,2\pi$, independently of the splitting $\Delta\epsilon$.
Large visibilities indeed demand a near perfect suppression of even or odd subparity dynamics \BrackEq{eq_subparity_decoupling} for \emph{both} dot levels separately, thus requiring $\Delta \phi \sim 0$. However, $\Delta\phi$ is not expected to be precisely tunable in practice, as it depends on the dot orbital details. Yet, even so, the intermediate-time readout offers a finite $(\epsilon,
\Delta\phi)$-regime with higher than stationary visibilities, especially at larger bias $\Delta\mu = 10\lambda$, see also appendix~\ref{app_further}.   

\begin{figure}[t!!]
\includegraphics[width=\linewidth]{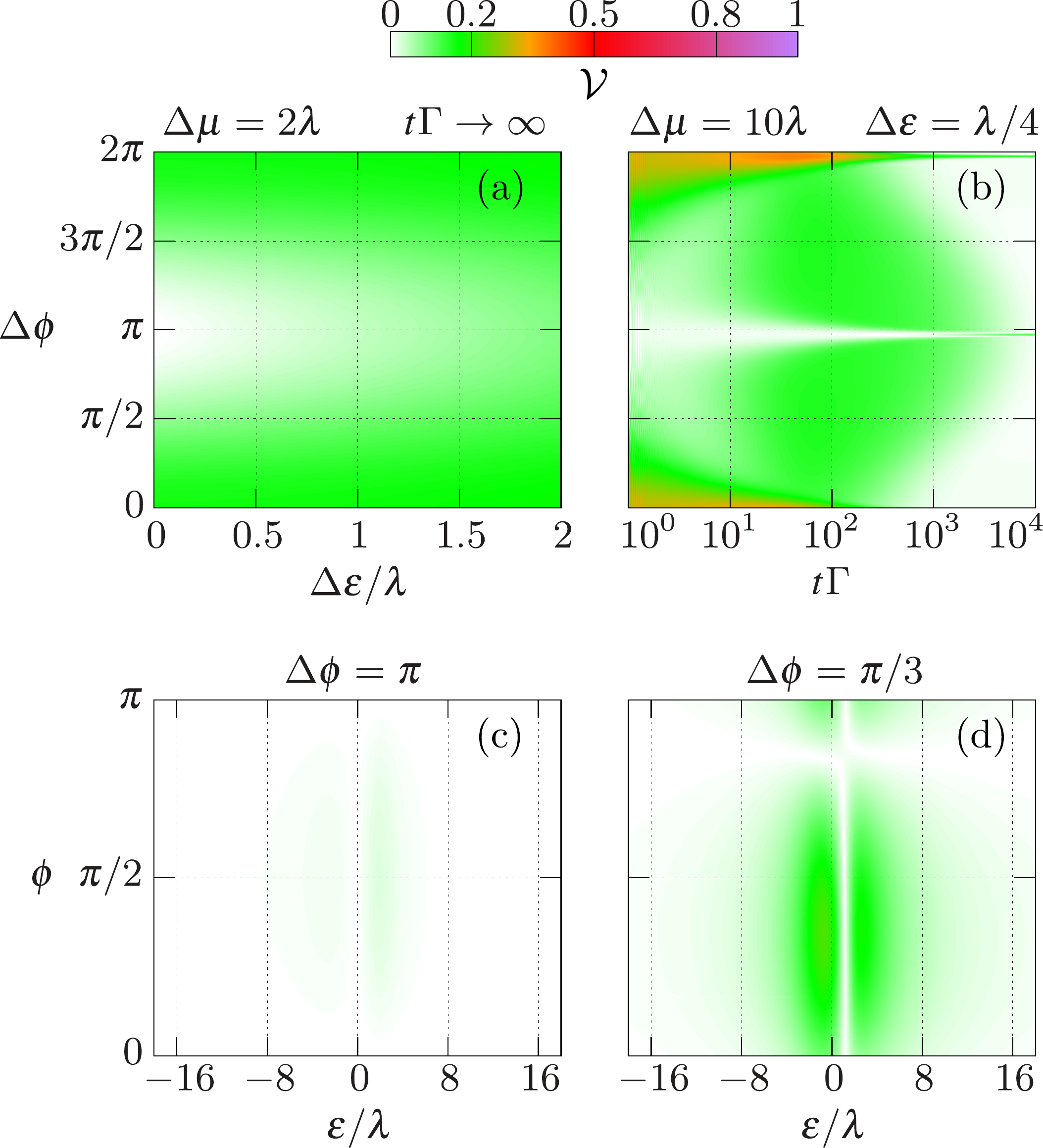}
\caption{(a) Visibility \eq{eq_visbility_time} in the stationary limit $t \rightarrow \infty$ for $M = 2$ dot levels as a function of level splitting $\Delta\epsilon = \epsilon_2 - \epsilon_1$ and relative tunneling phase $\Delta\phi$ \BrackEq{eq_tunnel_couplings}, given a constant average detuning $\epsilon = (\epsilon_1 + \epsilon_2)/2 = 2\lambda$ and reference phase $\phi = \pi/2$. (b) Visibility as a function of time $t\Gamma = tg\lambda$ and $\Delta\phi$ for an initially empty dot, $N(t = 0) = 0$, a detuning $\epsilon = 4\lambda$ and a reference phase $\phi = 1.05\pi/2$ deviating from the optimal case. (c,d) Stationary visibility with constant splitting $\Delta\epsilon = \lambda/4$ for two different $\Delta\phi$, plotted as a function of $\phi$  and $\epsilon$ at moderate bias $\Delta\mu = 2\lambda$. Parameters common to all panels are $g = 0.05$, $T = \lambda/5$, $\omega_c = 20\lambda$. \label{fig_visibility_double_phasedep}}
\end{figure}

To better estimate how much interference-related visibility loss may in practice occur for a given setup, it is crucial to know how small the dot level splitting $\Delta\epsilon$ can be before the effect kicks in. Figure~\ref{fig_visibility_double_phasedep}(a) therefore displays the stationary visibility $\mathcal{V}$ as a function of $\Delta\epsilon$ and $\Delta\phi$, fixing an optimal reference phase $\phi = \pi/2$ and a detuning $\epsilon = 4\lambda$ with sizable stationary visibility around $\Delta\phi = 0$ $[$\Fig{fig_visibility_double}(a)$]$. The plot clearly shows a considerable reduction for splittings $\Delta\epsilon \leq \lambda$ in a finite range $3\pi/4 < \Delta\phi < 5\pi/4$. For example, the visibility for splittings $\Delta\epsilon \approx \lambda/2$ is already nearly as suppressed as for $\Delta\epsilon = \lambda/4$. Hence, to avoid interference related visibility loss, stationary readout should be performed in a dot-energy range with splittings $\Delta\epsilon$ at least larger than the tunnel coupling $\lambda$.

With metastable, intermediate-time visibilities predicted to exceed $\mathcal{V}(t\rightarrow\infty)$ at $\phi = \pi/2$ even for two interfering dot levels $[$\Figs{fig_visibility_double}(c,d), and \Figs{fig_visibility_double_comp}(c,d) in appendix~\ref{app_further}$]$, it is interesting to see how long this metastability lasts with small, possibly noise related deviations from the optimal reference phase $\phi = \pi/2$ and Majorana-symmetric amplitudes $|\lambda_{2j}| = |\lambda_{1j}|$. We analyze this by setting $\phi = 1.05\pi/2$, which, roughly extrapolated from \Eq{eq_hamiltonian_single} and in agreement with numerical checks, would be similar to a situation with amplitude asymmetry $|\lambda_{2j}| \approx 0.95|\lambda_{1j}|$ at optimal phase $\phi = \pi/2$. We plot the resulting visibility $\mathcal{V}(t,\Delta\phi)$ in \Fig{fig_visibility_double_phasedep}(b) with small constant splitting $\Delta\epsilon = \lambda/4$, large bias $\Delta\mu = 10\lambda$ and detuning $\epsilon = 4\lambda$. After a decay from an initially empty dot during times $t\Gamma = tg\lambda \leq 10$, we find a finite visibility plateau up to $t\Gamma \sim 2000$; strong enhancements at nearly optimal relative phase $\Delta\phi \rightarrow 0$ are observed until $t\Gamma \sim 200$. These times would correspond to $t \sim 5\,\mu\text{s}$ and, respectively, $t \sim 0.5\,\mu\text{s}$ for the example of $\lambda \sim 10\,\mu\text{eV} \sim 2\,\text{GHz}$ and $g = 0.05$. While a temporal resolution $t \ll \mu\text{s}$ may be challenging, we expect time spans well above $1\,\mu\text{s}$ to be within experimentally achievable limits. 
Despite the practical challenges, this suggests intermediate-time readout as proposed in \Sec{sec2e} as a viable option if the dot-Majorana tunnel couplings can be tuned sufficiently well, even if multi-level effects enter. Strong visibility enhancements as in the optimal single-level case are unlikely to be observable, but the charge visibility at optimal $\phi$ is expected to be as large or even much larger than in stationary readout, especially at larger biases $\Delta\mu \gg \lambda$.

Finally, it is of practical interest whether the stationary visibility loss for small splittings $\Delta\epsilon < \lambda$ and relative phases $\Delta\phi \approx \pi$ can be mitigated by retuning the reference phase \emph{away from} $\phi = \pi/2$ via an out-of-plane flux. Figures \ref{fig_visibility_double_phasedep}(c,d) examine this by comparing the $(\epsilon,\phi)$-dependent stationary visibility at $\Delta\phi = \pi$ to the one at $\Delta\phi = \pi/3$. It confirms that the choice $\phi = \pi/2$ is indeed close to the optimum. Thus, at small splittings, the visibility loss due to a constant relative tunneling phase $\Delta\phi \approx \pi$ cannot be simply mitigated by retuning $\phi$ via, e.g., adjusting the magnetic flux. Conversely, for $\Delta\phi = \pi/3$, near optimal visibility does not critically depend on $\phi$ in a sizable range.

\subsection{Physical origin of state-dependent coupling phase}
\label{sec3b}
In the following, we investigate in a more realistic model the state-dependent coupling phases $\Delta\phi$ and splittings $\Delta\epsilon$. By better understanding how these parameters are physically determined in a concrete system, we aim to find out how likely the above described, detrimental condition $\Delta\phi \approx \pi$ with level spacings $\Delta\epsilon < \lambda$ is. 

It is intuitively clear from the well-known particle-in-a-box problem that reducing the lateral dimensions of the dot increases the typical level splitting $\Delta\epsilon$. For, e. g., $\lambda \sim \text{GHz}$ as suggested above, practically achievable lateral dimensions $\lesssim 100\,\text{nm}$ would in most cases guarantee $\Delta\epsilon > \lambda$ at reasonable fillings. However, this straightforward approach to avoid spacings $\Delta\epsilon < \lambda$ may not always be preferable in experiment. Already the box qubit setup studied in this work requires sufficient distance between the superconducting wires for the Majoranas to not hybridize directly with each other, see \Fig{fig_setup}. The dot thus needs to be similarly long in order to couple to both wires with sizable amplitude. Moreover, one large dot with tunable couplings to selectively read out several separate Majorana pairs may result in lower device complexity compared to several smaller readout dots for each individual Majorana pair. It is thus experimentally relevant to also consider dot sizes in which splittings $\Delta\epsilon < \lambda \sim \text{GHz}$ cannot be ruled out beforehand. 

\begin{figure}[t!!]
\includegraphics[width=\linewidth]{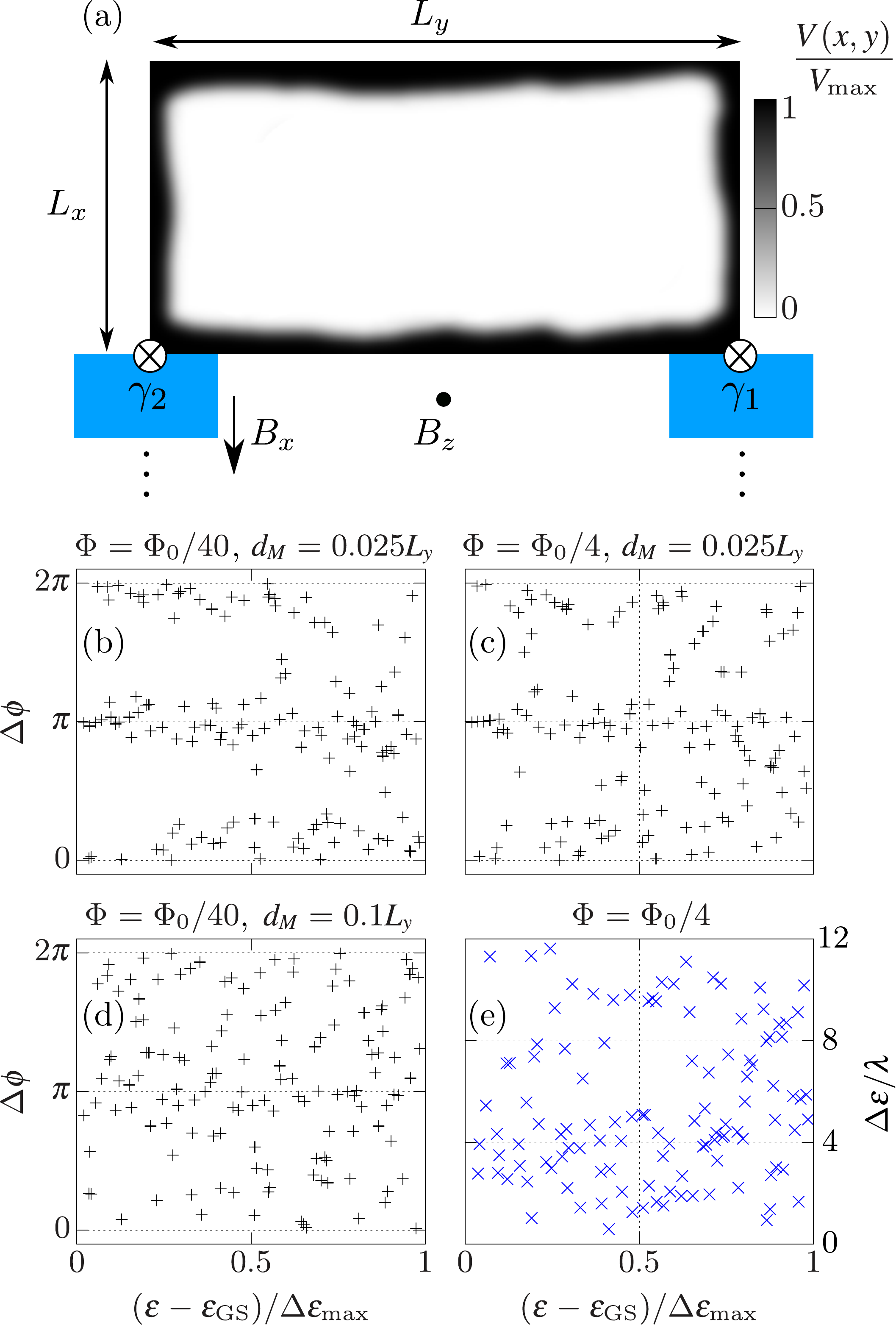}
\caption{(a) A 2D quantum well (central rectangle) with lateral dimensions $L_y = 2L_x = 800\,\text{nm}$ coupling to Majorana modes, marked by crosses at the end of two topological nanowires (blue). 
The color scale shows the potential landscape $V(x,y)$ of the well up to $V_{\text{max}} = \pi^2/(2m^*(10l)^2)$, corresponding to a wavelength at the lower end of what can be resolved reliably with a grid spacing $l$.  
To mimick an infinitely high potential wall and avoid numerical issues, points at $V(x,y) = V_{\text{max}}$ are entirely excluded from the dynamics, see appendix \ref{app_qd}.\\
(b-e) Relative phase shifts $\Delta\phi$ \BrackEq{eq_phase_shifts} and energy splittings $\Delta\epsilon$ \BrackEq{eq_splittings} between two neighboring states $j-1,j$ ($\Delta\epsilon > 12\lambda$ not shown for better readability). The samples are ordered according to energy $\epsilon$ relative to the ground level $\epsilon_{\text{GS}}$. The scale $\Delta\epsilon_{\text{max}} = \epsilon_{141} - \epsilon_{\text{GS}}$ is the difference between ground state and highest energy of the first $141$ levels. The typically expected tunneling amplitude is set to $\lambda = 10\,\mu\text{eV} \approx 2.4\,\text{GHz}$. Rashba spin-orbit coupling $\alpha = 20\,\text{meVnm}$, g-factor $g_{\text{L}} = 14$ and effective mass $m^* = 0.026m_e$ in the quantum well are consistent with bulk InAs, and a strong Zeeman field $B_x = 1000\lambda$ along the x direction is included next to a flux $\Phi \sim B_z$ in z direction perpendicular to dot-Majorana plane. We discretize the quantum well onto $230\times 460$ grid points.
}
\label{fig_phaseshift_splitting}
\end{figure}

More specifically, we study phase differences and level splittings for the dot-Majorana model system sketched in \Fig{fig_phaseshift_splitting}(a). The two-dimensional dot wave functions $\Psi_j(\mathbf{x})$ and their energies $\epsilon_j$ are obtained by modelling the dot as an approximately rectangular potential well with a flat potential in the bulk and slightly irregular boundary, see large rectangle in \Fig{fig_phaseshift_splitting}(a). Here, we focus on asymmetric potential landscape dimensions $L_y = 2L_x = 800\,\text{nm}$ representing large elongated islands stretching between the ends of two parallel wires (blue rectangles), see also \Fig{fig_setup}. The corresponding Schr\"odinger equation is dicretized onto a 2d nearest-neighbor-hopping system with discretization length $l$, as detailed in appendix~\ref{app_qd}. We partly diagonalize this system with the Lanczos method, limited to kinetic energies equivalent to wavelengths an order of magnitude larger than the discretization length. The effective mass $m^* = 0.026m_e$ in terms of free electron mass $m_e$, g-factor $g_{\text{L}} = 14$, and the spin-orbit coupling strength $\alpha = 20\,\text{meVnm}$ are set to typically measured values in bulk InAs systems~\cite{Cardona1961Feb,Vurgaftman2001Jun,Liang2012Jun,O'Farrell2018Dec,Kiselev1998Dec}. Furthermore, next to the large, spin-polarizing Zeeman energy field $B_x$ parallel to the nanowires, a perpendicular magnetic flux $\Phi = 2m_e L_yL_xB_z/g_{\text{L}}$ interacts both with the spin and orbital degree of freedom of the dot electrons. The effect of Coulomb repulsion is simplified as an occupation-dependent energy shift that leaves the single-particle states invariant. This commonly used \emph{constant interaction approximation} thus qualitatively shows how wavenumber, spin-orbit coupling and the externally tunable flux affect the energies $\epsilon_j$, nearest-neighbor splittings $\Delta\epsilon_j = \epsilon_j - \epsilon_{j-1}$ and wavefunctions $\Psi_j(\mathbf{x})$ for the low-lying dot orbitals $j$.

The tunnel coupling phases between the dot states and the Majoranas $\gamma_{1,2}$ are then extracted from the wavefunction overlaps at the dot-Majorana boundaries. We assume that the strong Zeeman field $B_x$ along the topological wires spin-polarizes the $\gamma_{1,2}$ into the negative x direction at the wire ends. The corresponding wave functions are then simplified as exponential tails in the dot-tetron plane, with decay length $d_{M}$ extending into the tunnel barriers at the boundaries of the quantum dot potential. The electron-like component reads
\begin{equation}
 \gamma_i(\mathbf{x}) \equiv K\begin{pmatrix}1 \\ -1\end{pmatrix}e^{-|\mathbf{x}-\mathbf{x}_i|/d_M}.\label{eq_wave_majorana}
\end{equation}
The positions $\mathbf{x}_i$ are the reference positions at the dot boundary corners, marked by the crosses in \Fig{fig_phaseshift_splitting}(a), at which the corresponding Majorana wave functions $\gamma_i$ assume the reference value $K$. The spin in \Eq{eq_wave_majorana} is represented in the spin-z eigenbasis. Finally, with both dot and Majorana wave functions at hand, we estimate the tunneling phases \eq{eq_phase_asymmetry} between Majorana $\gamma_2$ and $\gamma_1$ by
\begin{equation}
 \phi_{2j} - \phi_{1j} = \text{arg}\sqbrack{\sum_{\mathbf{x}}\gamma_2(\mathbf{x})\Psi_j(\mathbf{x})} - \text{arg}\sqbrack{\sum_{\mathbf{x}}\gamma_1(\mathbf{x})\Psi_j(\mathbf{x})}\label{eq_phase_asymmetry_physical}.
\end{equation}
The sum $\sum_{\mathbf{x}}$ over all discrete dot lattice sites approximates the overlap integral. 

Using the above described diagonalization procedure and \Eq{eq_phase_asymmetry_physical}, we can calculate the nearest-neighbor splittings and relative phase shifts
\begin{gather}
\Delta\epsilon_j = \epsilon_j - \epsilon_{j-1}\label{eq_splittings}\\
 \Delta\phi_j = (\phi_{2j} - \phi_{1j}) - (\phi_{2j-1} - \phi_{1j-1})\label{eq_phase_shifts}.
\end{gather}
Note that our approach does not provide the absolute tunneling amplitudes $|\lambda_{ij}|$ for each individual Majorana $i$ and dot level $j$: unlike the relative phases \eq{eq_phase_shifts}, these also depend on the Majorana wave function amplitude $\sim K$ which cannot be obtained from our simplified model. Rather, we fix the tunneling frequencies to typically expected or desired values in experiments, $|\lambda_{ij}| \sim \lambda = 10\,\mu\text{eV} \sim 2\,\text{GHz}$, and use $\lambda$ as the reference energy scale for better comparison to the low-energy model \eq{eq_hamiltonian}.
With this assumption, the splittings \eq{eq_splittings} and relative phases \eq{eq_phase_shifts} are shown in \Fig{fig_phaseshift_splitting}(b-e) as a function of energy difference $\epsilon - \epsilon_{\text{GS}}$ between the lower level $\epsilon_{j-1}$ of each splitting and the ground state energy $\epsilon_{\text{GS}}$. We consider different combinations of Majorana decay lengths $d_L$ and perpendicular magnetic flux $\Phi$.
At small flux $\Phi = \Phi_0/40$ and decay length $d_M = 0.025L_y$, the phase differences are bi-modally distributed around $\Delta\phi = 0$ (modulo $2\pi$) and $\Delta\phi = \pi$ at low energy $\epsilon \gtrsim \epsilon_{\text{GS}}$, see \Fig{fig_phaseshift_splitting}(b). This reflects the typical particle-in-a-box behavior in which wave functions with subsequent wavenumbers have opposite relative signs close to the boundary.
For larger energies and hence larger wavenumbers, the phase distribution spreads out due to stronger spin-orbit coupling and stronger coupling to the perpendicular flux. At $\Phi = \Phi_0/4$ considered in \Fig{fig_phaseshift_splitting}(c), the phase spread therefore also becomes sizable at lower energies. When increasing the decay length $d_M$ as in \Fig{fig_phaseshift_splitting}(d), the distribution further broadens as the overlaps \eq{eq_phase_asymmetry_physical} become more susceptible to lower-frequency spatial variations.

As shown by the histogram in \Fig{fig_hist} of appendix~\ref{app_qd}, the distribution of the energy splittings $\Delta \epsilon$ in \Fig{fig_phaseshift_splitting}(e) displays two main peaks at low energy. The first is located around $\Delta\epsilon \approx 40 \mu{\text eV} \sim 4\lambda$, the second at $\Delta\epsilon \approx 100 \mu{\text eV} \sim 10 \lambda$.  We interpret the former as the result of the random confining potential, and the latter as the result of the systematic features of a regular rectangular quantum dot with spin-orbit coupling. In particular, the splitting distribution can be roughly approximated by the overlap of a Wigner surmise obtained from a random Gaussian unitary ensemble \cite{Guhr1998Jun,Alhassid2000Oct,Mehta2004Oct}  $\sim (\Delta\epsilon/\langle\Delta\epsilon\rangle)^2e^{-\frac{4}{\pi}\nbrack{\Delta\epsilon/\langle\Delta\epsilon\rangle}^2}$ with average $\left\langle \Delta \epsilon \right\rangle \approx 45 \mu{\text eV}$ and the additional peak located at $10 \lambda$ (comparable with $\pi \alpha/L_y$), which shifts the overall average spacing at $\langle\Delta\epsilon\rangle \approx 70 \mu{\text eV}$. 

This implies that, due to the random potential energy level repulsion, small $\Delta\epsilon \lesssim \lambda$ are not encountered as regularly as in a clean symmetric system with similar dimensions and effective mass. Nevertheless, \Fig{fig_phaseshift_splitting}(e) still exhibits a finite number of $\Delta\epsilon \lesssim 10 \mu{\text eV} \sim \lambda$ across the entire energy range, including low filling numbers close to $\epsilon = \epsilon_{\text{GS}}$.  Furthermore, we point out that $\lambda = 10 \mu{\text eV}$ is only a \textit{typical estimate} of the tunnel coupling strength; hence, the splitting distribution could shift to even smaller ratios $\Delta \epsilon/\lambda$, depending on the details of the tunnel barriers.

As pointed out in \Sec{sec3a}, splittings $\Delta \epsilon$ below the tunnel coupling $\lambda$ could cause relevant visibility reductions in parity-to-charge conversion. Combining the above findings, we conclude that, in general, this visibility loss cannot be ruled out beforehand in experiments relying on large quantum dots or islands. For the parameters chosen here, the condition $\Delta\phi \approx \pi$ is regularly encountered at low dot occupations. More than one subsequent splitting $\Delta\epsilon < \lambda$, i. e. the case $M > 2$ in \Eq{eq_hamiltonian} seems rare in dots or islands with irregular spatial features, suggesting that the visibility loss can be mitigated by bringing a different dot state in resonance with the Majoranas. However, the possibility of having to do so needs to be accounted for, and the situation becomes more difficult with larger coupling $\lambda$, or with larger dot size or effective mass at constant coupling. Also, the splittings might be distributed within a narrower range in more symmetric systems, so that a simple gate voltage shift may no longer suffice. In the following final section~\ref{sec3c}, we therefore also estimate the potential effect of multi-level interference in such cases, including more than $2$ dot levels with splittings $\Delta\epsilon \lesssim \lambda$.

\subsection{Average visibility in many-state dot}
\label{sec3c}
Let us finish by comparing the stationary visibility in the case of $M = 2$ dot orbitals $j$ to the case $M = 8$. To be less specific to the concrete parameters, we analyze sample averages $\langle|\mathcal{V}(t\rightarrow\infty)|\rangle_{\Delta\epsilon,\Delta\phi}$ over splittings $\Delta\epsilon$ and orbital-dependent relative phases $\Delta\phi$.

More precisely, we take the local Hamiltonian $H$ \BrackEq{eq_hamiltonian} and the energy-dependent jump operator $L$ \BrackEq{eq_jump_operator} with  
\begin{equation}
 \hat{\epsilon}_{1} = \epsilon \quad,\quad \hat{\epsilon}_{j>1} = \hat{\epsilon}_{j-1} + \Delta\epsilon + \widehat{\delta\epsilon}\label{eq_level_pos_random}
\end{equation}
and tunnel couplings
\begin{equation}
 \lambda_{1j} = \lambda \quad,\quad \lambda_{2j} = \lambda e^{i\widehat{\phi}_j}\label{eq_tunnel_coupling_random}
\end{equation}
with random phases
\begin{equation}
 \widehat{\phi}_{j=1} = \phi \quad,\quad \widehat{\phi}_{j>1} = \widehat{\phi}_{j-1} + \widehat{\Delta\phi} + \widehat{\delta\phi}.\label{eq_phase_diff_random}
\end{equation}
By inserting \Eq{eq_level_pos_random} into \eq{eq_splittings}, each level splitting is therefore given by a constant deterministic splitting $\Delta\epsilon$ plus a random level spread $\widehat{\delta\epsilon}$. We approximate $\widehat{\delta\epsilon}$ as uniformly distributed in the interval $[-\delta\epsilon,\delta\epsilon]$ with $0 \leq 2\delta\epsilon < \Delta\epsilon$. This forces a non-zero minimum splitting, assuming that the systems is subject to some form of level repulsion that prohibits perfect degeneracy. The corresponding phase shifts $\Delta\phi_j$ defined in \Eq{eq_phase_shifts} consist, according to \Eq{eq_phase_diff_random}, of two random contributions. Sampled from $\{0,\pi\}$, $\widehat{\Delta\phi}$ reflects the bi-modal phase distribution observed in \Fig{fig_phaseshift_splitting}(b). The term $\widehat{\delta\phi}$ adds a Gaussian spread around these phase differences with mean $0$ and standard deviation $\sigma = \delta\phi$.

\begin{figure}[t!!]
\includegraphics[width=\linewidth]{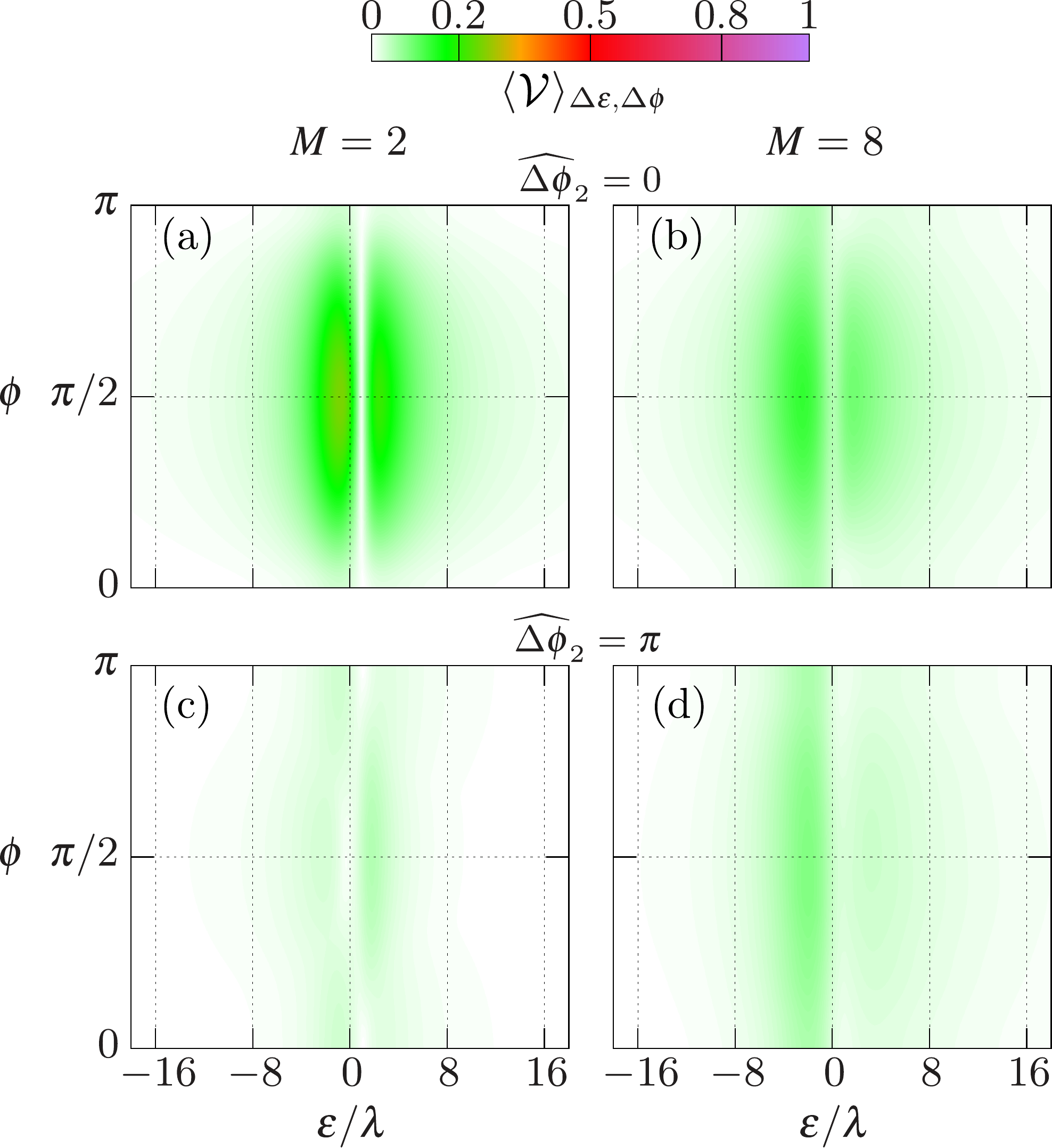}
\caption{The $(\epsilon,\phi)$-dependence of the visibility \eq{eq_visbility_time} in the stationary limit $t\rightarrow\infty$, averaged over $S = 100$ samples with $M = 2$ (a,c) and $M = 8$ (b,d) dot levels $j$. Random splittings $\widehat{\Delta\epsilon}_j$ and coupling phase differences $\widehat{\Delta\phi}_j$ are chosen according to \Eqs{eq_level_pos_random},\eq{eq_tunnel_coupling_random},\eq{eq_phase_diff_random} and the description thereafter. Parameters are $\Delta\epsilon = \lambda/2$, $\delta\epsilon = \lambda/4$, $\delta\phi = \pi/8$, $g = 0.05$, $T = \lambda/5$, $\Delta\mu = 2\lambda$, $\omega_c = 20\lambda$. \label{fig_multi_level}}
\end{figure}

Between the second and first level, we fix the binary phase contribution $\widehat{\Delta\phi}_2$  to the same constant value in each random sample of $H$. This highlights how additional levels $M > 2$ affect both the worst case scenario $\widehat{\Delta\phi}_2 = \pi$, and the optimal case in which the system can be tuned to $\widehat{\Delta\phi}_2 = 0$. The left and right panels in \Fig{fig_multi_level} show, respectively, the sample-averaged visibility for $M = 2$ and $M = 8$ as a function of reference phase $\phi$ and lowest level $\epsilon$ for $S = 100$ samples. 
The result for $M = 2$ in the optimal case $\widehat{\Delta\phi}_2 = 0$ in \Fig{fig_multi_level}(a) is essentially an averaged version of \Fig{fig_visibility_double_phasedep}(d): peak visibilities exceed $1/4$ around $\phi = \pi/2$ for levels around $\epsilon = 0$, with a slightly larger visibility for $\epsilon < 0$ compared to $\epsilon > 0$.  Increasing to $M = 8$ levels, \Fig{fig_multi_level}(b) exhibits a more smeared out function with both a small loss in peak average visibility and a more pronounced level-asymmetry around $\epsilon = 0$.
The worst case $\widehat{\Delta\phi}_2 = \pi$ for $M = 2$ levels plotted in \Fig{fig_multi_level}(c) is similar to the situation \Fig{fig_visibility_double_phasedep}(c). Hence, for only small phase spread $\delta\phi = \pi/8$ around $\Delta\phi = \pi$ and splittings distributed below $\Delta\epsilon = \lambda$, the visibility remains strongly suppressed. For $M = 8$ levels, \Fig{fig_multi_level}(c) shows a more pronounced asymmetry around $\epsilon \approx 0$ that moderately improves peak average visibility over $M = 2$. 

On average, the interference of many levels seems to have an additional smoothing effect on the visibilities, i. e., peaks shrink but strongly suppressed visibilities are also partly restored, in particular for the lowest level around $\epsilon = 0$. The latter is to a considerable degree a result of stronger $\epsilon$-asymmetry. The crucial point is, on an intuitive level, that while 10 levels give rise to 5 times more possibilities of singly occupying the dot than 2 levels, each case still has only one many-body state with an empty dot in each subparity sector. Any $s$-dependence in the dot-Majorana coupling, and hence the even-odd visibility therefore tends to be amplified for $\epsilon < 0$ with a predominantly occupied dot compared to $\epsilon > 0$ with a preferably empty dot.

\section{Conclusion}
\label{sec5}
We have theoretically analyzed how parity-to-charge conversion of a Majorana box qubit with a quantum dot and a QPC charge sensor is affected by the interference of multiple dot levels. We find that if two single-particle dot states couple with comparable strength but roughly $\pi$-shifted relative phase to the Majorana modes, the net visibility gets significantly reduced for dot-level splittings smaller than the tunnel coupling strength. These $\pi$-shifts can often occur in dots at low charge filling and, in general, cannot be simply avoided by an appropriate external magnetic flux. The resulting visibility loss may therefore become relevant in, e.g., large dots with typical splittings $\Delta\epsilon\lesssim 10\,\mu\text{eV}$ and tunneling frequencies tuned towards $\lambda\sim\text{GHz}$. However, we show that by performing an \emph{intermediate-time} readout protocol, the interference-related visibility loss can partly be mitigated. Such a protocol presents the additional advantage of yielding much less reduced visibilities for QPC bias potentials exceeding the induced qubit splitting. Finally, multi-level interference is predicted to average out in realistic systems when more than two dot orbitals couple with similar strength, thus mitigating further the visibility loss which characterizes the analogous two-level case with a $\pi$-shift of the tunneling phase. 
With larger dots or Coulomb islands being common building blocks in the here addressed type of devices, our results highlight important limitations and mitigation strategies for future experiments on quantum-dot based parity-to-charge conversion.

\begin{acknowledgments}
We thank Morten Munk, Reinhold Egger, Eoin O'Farrel, Felix Passmann, Serwan Asaad, Charles Marcus, Max Geier, Frederik Nathan and Mark Rudner for valuable discussions. This project has received funding from the European Research Council (ERC) under the European Union’s Horizon 2020 research and innovation programme under Grant Agreement No. 856526. This research was supported by the Danish National Research Foundation, the Danish Council for Independent Research $\vert$ Natural Sciences, the Swedish Research Council (VR), and by the Microsoft Corporation. M.B. is supported by the Villum Foundation (Research Grant No. 25310). 
\end{acknowledgments}

\appendix

\section{Lamb shift}
\label{app_lamb}

\begin{figure}[t!!]
\includegraphics[width=\linewidth]{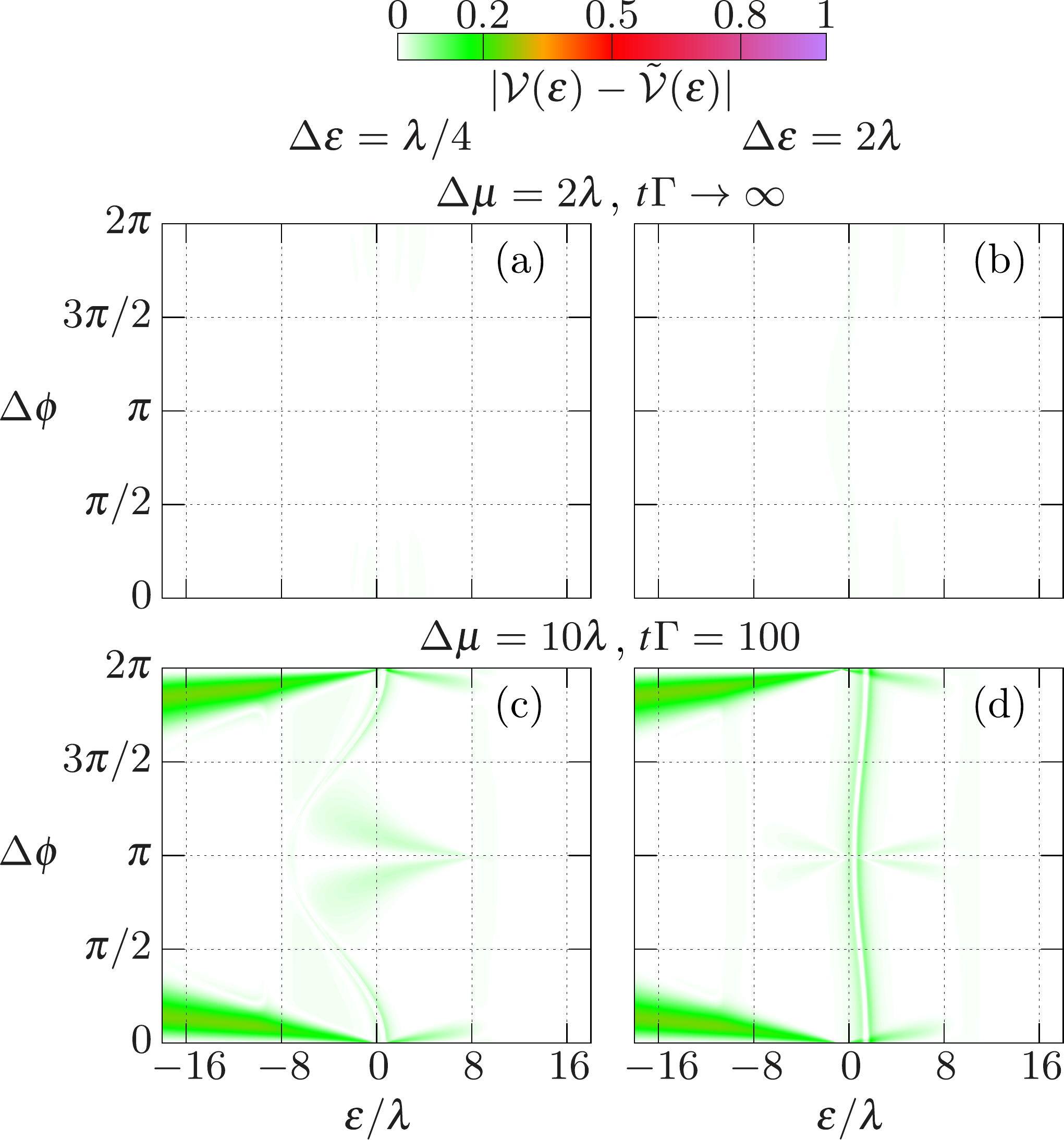}
\caption{Deviation of visibility with Lamb shift $\mathcal{V}(\epsilon)$ to level-shifted visibility without Lamb shift $\tilde{\mathcal{V}}(\epsilon) = \mathcal{V}(\epsilon - \lambda/2,
\Lambda = 0)$, as a function of average detuning $\epsilon = (\epsilon_1 + \epsilon_2)/2$ and relative tunneling phase $\Delta\phi$ \BrackEq{eq_tunnel_couplings} for $M = 2$ levels. Parameters and initial dot state are the same as in \Figs{fig_visibility_double}(a-d). \label{fig_lamb}}
\end{figure}

The Lamb shift $\Lambda$ affecting the local coherent dynamics in the quantum master equation \eq{eq_master} is obtained from \Refe{Nathan2020Sep}:
\begin{gather}
 \Lambda = \sum_{\eta,\eta',\tilde{\eta}}f(E_{\eta} - E_{\tilde{\eta}},E_{\tilde{\eta}} - E_{\eta'})\bra{\eta}N\ket{\tilde{\eta}}\bra{\tilde{\eta}}N\ket{\eta'} \times \ket{\eta}\bra{\eta'}\notag\\
 f(E,E') = \mathcal{P}\int_{-\infty}^\infty d\omega\frac{-\sqrt{\BQPC(\omega + E)}\sqrt{\BQPC(\omega - E')}}{2\pi\omega},\label{eq_lamb}
\end{gather}
where $\ket{\eta}$ and $E_{\eta}$ are the manybody eigenstates and corresponding energies of the Hamiltonian \eq{eq_hamiltonian}, $\BQPC(\omega)$ is the QPC spectral function \eq{eq_qpc_bath_correlation_fourier_final}, and $\mathcal{P}$ indicates the Cauchy principal value.
In the low temperature limit, \Refe{Munk2020Aug} has reduced the integral $f(E,E')$ to a combination of exponential integrals $\text{Ei}(x) = \int_{-\infty}^xdx' e^{x'}/x'$ in the special case of only a single dot level, $M = 1$. The corresponding analysis has shown that for a spectral cutoff frequency $\omega_c$  \BrackEq{eq_qpc_bath_correlation_fourier_final} larger than the tunneling induced qubit splitting and the QPC potential bias, $\omega_c \gg \lambda,\Delta\mu$, the Lamb shift $\Lambda$ reduces to a subparity$(s)$-independent shift of the dot level $\epsilon$. For the here relevant multi-level case $M > 1$, we compute the integral $f(E,E')$ numerically over the frequency intervals $[-\Delta\omega,0)$ and $(0,\Delta\omega]$ using Simpson's $1/3$ rule:
\begin{align}
 f(E,E') &\approx -\frac{1}{6\pi}\sum_{n_\omega = 1}^{N_\omega\text{ odd}}\frac{c(n_\omega)}{n_\omega}\tilde{f}\left(\frac{n_\omega}{N_\omega}\Delta\omega,E,E'\right)\notag\\
 \tilde{f}\left(\omega,E,E'\right) &= \sqrt{\BQPC(\omega + E)}\sqrt{\BQPC(\omega - E')}\notag\\
 &\phantom{=}- \sqrt{\BQPC(-\omega + E)}\sqrt{\BQPC(-\omega - E')}\notag\\
 c(n_\omega) &= \begin{cases} 1 & n_\omega = 1, N_\omega \\ 4 - 2(n_\omega\text{ mod }2) & 1 < n_\omega < N_\omega\end{cases}.\label{eq_lamb_func}
\end{align}
The integration boundary $\Delta\omega = 10\omega_c$ used for all plots in this paper is large enough for the spectral function $\BQPC(\omega) \sim e^{-|\omega|/\omega_c}$ \BrackEq{eq_qpc_bath_correlation_fourier_final} to have decayed. The number of integration points $N_\omega = 201$ chosen in each case is found to yield acceptable convergence. In generating the data for \Fig{fig_multi_level}, we avoid prohibitive numerical costs by precalculating $f(E,E')$ according to \Eq{eq_lamb_func} on a discrete $(E,E')$-lattice with sufficient resolution and energy range. The evaluation of \Eq{eq_lamb} then samples $f(E,E')$ from this lattice with bilinear interpolation.

\begin{figure}[t!!]
\includegraphics[width=\linewidth]{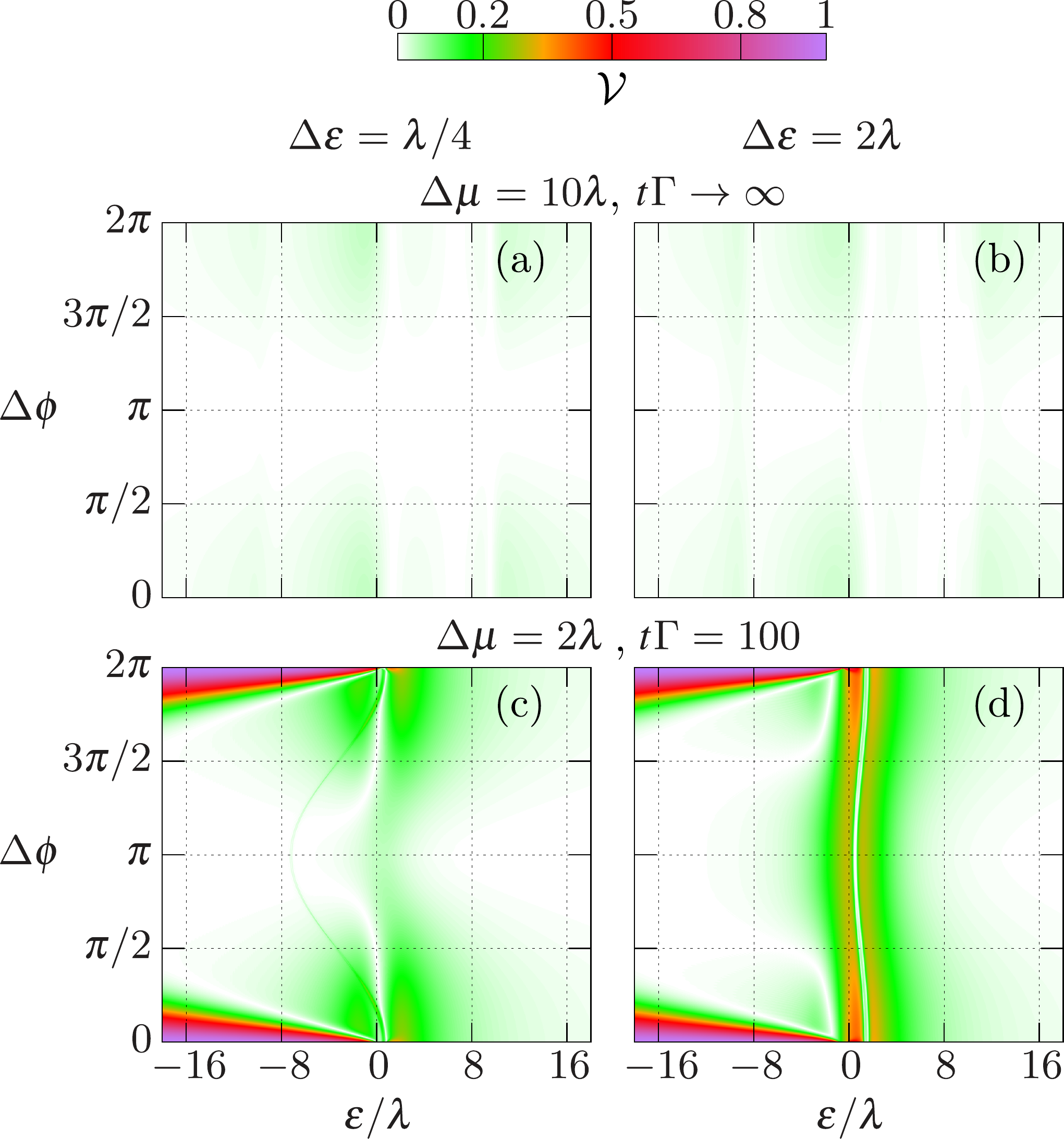}
\caption{Charge visibility \eq{eq_visbility_time} as a function of relative tunneling phase $\Delta\phi$ \BrackEq{eq_tunnel_couplings} and average detuning $\epsilon = (\epsilon_1 + \epsilon_2)/2$ of $M = 2$ dot levels, with biases $\Delta\mu$ complementary to those chosen in \Figs{fig_visibility_double}(a-d). In (c,d), we assume $N(t = 0) = 0$. In all panels, we set $\phi = \pi/2$, $g = 0.05$, $T = \lambda/5$, $\omega_c = 20\lambda$. \label{fig_visibility_double_comp}}
\end{figure}

As an example of how the Lamb shift affects the visibility, \Fig{fig_lamb} compares $\mathcal{V}$ in the case of $M = 2$ dot levels and a finite $\Lambda$ to the visibility with the Lamb shift set to $\Lambda = 0$. In the stationary limit, one finds that $\mathcal{V}(\epsilon - \lambda/2,\Lambda = 0)$, i. e., evaluated at the slightly shifted detuning $\epsilon - \lambda/2$ indeed closely matches $\mathcal{V}(\epsilon)$, just as in the case $M = 1$ studied in \Refe{Munk2020Aug}.
However, in the correctly Lamb-shifted visibility $\mathcal{V}(\epsilon)$ at the intermediate time $t\Gamma = 100$ and optimal reference phase $\phi = \pi/2$, the mere $\epsilon$-shift is accompanied by a larger $\Delta\phi$ range around $\Delta\phi = 0,2\pi$ with enhanced metastable visibility; the interval around $\Delta\phi = \pi$ with suppressed visibility is likewise increased in $\mathcal{V}(\epsilon)$ compared to $\tilde{\mathcal{V}}(\epsilon)$. Hence, while not affecting the visibility \emph{qualitatively}, the quantitative effect of the Lamb shift is not negligible in the intermediate-time, metastable regime. It is therefore fully accounted for in our calculations.

\section{Further results for $M = 2$ dot levels}
\label{app_further}

In this short appendix, we provide further results to complement \Fig{fig_visibility_double} for the case of $M = 2$ dot levels. More specifically, \Fig{fig_visibility_double_comp} shows the $(\epsilon,\Delta\phi)$-dependence of the stationary visibility at a larger bias $\Delta\mu = 10\lambda$, and the intermediate-time visibility at a smaller bias $\Delta\mu = 2\lambda$. Figures~\ref{fig_visibility_double_comp}(a,b) confirm the bias-related stationary visibility loss already observed in \Fig{fig_visibility_single}(c) for the single-level case $(M = 1)$ at $\Delta\mu = 10\lambda$. Compared to the intermediate-time results from \Figs{fig_visibility_double}(c,d) at $\Delta\mu = 10\lambda$, the results for $\Delta\mu = 2\lambda$ in \Figs{fig_visibility_double_comp}(c,d) exhibit notably smaller $(\epsilon,\Delta\phi)$-regimes with moderate metastable visibility enhancements (green areas). This suggests that for intermediate-time readout, larger biases may in fact be preferable. 

\section{Quantum dot Hamiltonian}
\label{app_qd}

This appendix details how the two-dimensional quantum well discussed in \Sec{sec3c} and shown in \Fig{fig_phaseshift_splitting}(a) is modelled.
The continuous form of the considered single-particle Hamiltonian confined to the potential landscape in the $x,y$ plane reads $(|e| = \hbar = 1)$
\begin{align}
 H_{\text{dot}} &= \frac{\sqbrack{i\partial_x - A_x(x,y)}^2 + \sqbrack{i\partial_y - A_y(x,y)}^2}{2m^*} + V(x,y)\notag\\
 &+ \frac{1}{2}\vec{B}\cdot\vec{\sigma} + \alpha\sqbrack{\sigma_x(-i\partial_y) - \sigma_y(-i\partial_x)}.\label{eq_hamiltonian_well_continuous}
\end{align}
The first line contains the kinetic term in the presence of a magnetic field and the confining potential $V(x,y)$. 
We included a magnetic field $\sim B_x$ parallel to the potential well plane along the x-axis, and a field $\sim B_z$ perpendicular to the plane in z-direction. The corresponding vector potential is gauged to $\vec{A} = \frac{m_e}{g_{\text{L}}}(-B_z y,B_z x,2 B_x y)$, noting that $B_{x,z}$ are are Zeeman energies that yield the corresponding magnetic fields when divided by the Bohr magneton $1/(2m_e)$ and the g-factor $g_{\text{L}}$.
The potential energy landscape $V(x,y)$ is defined in \Fig{fig_phaseshift_splitting}(a) of the main text.
The second line in \Eq{eq_hamiltonian_well_continuous} adds the spin-Zeeman energy $\vec{B}\cdot\vec{\sigma} = B_x\sigma_x + B_z\sigma_z$ and the Rashba spin-orbit coupling $\alpha(\vec{\sigma}\times\vec{p})_z$. The latter equals the z-component of the cross product between Pauli vector $\vec{\sigma} = (\sigma_x,\sigma_y,\sigma_z)^T$ and momentum vector $\vec{p} = -i(\partial_x,\partial_y,\partial_z)^T$.

\begin{figure}
\includegraphics[width=\linewidth]{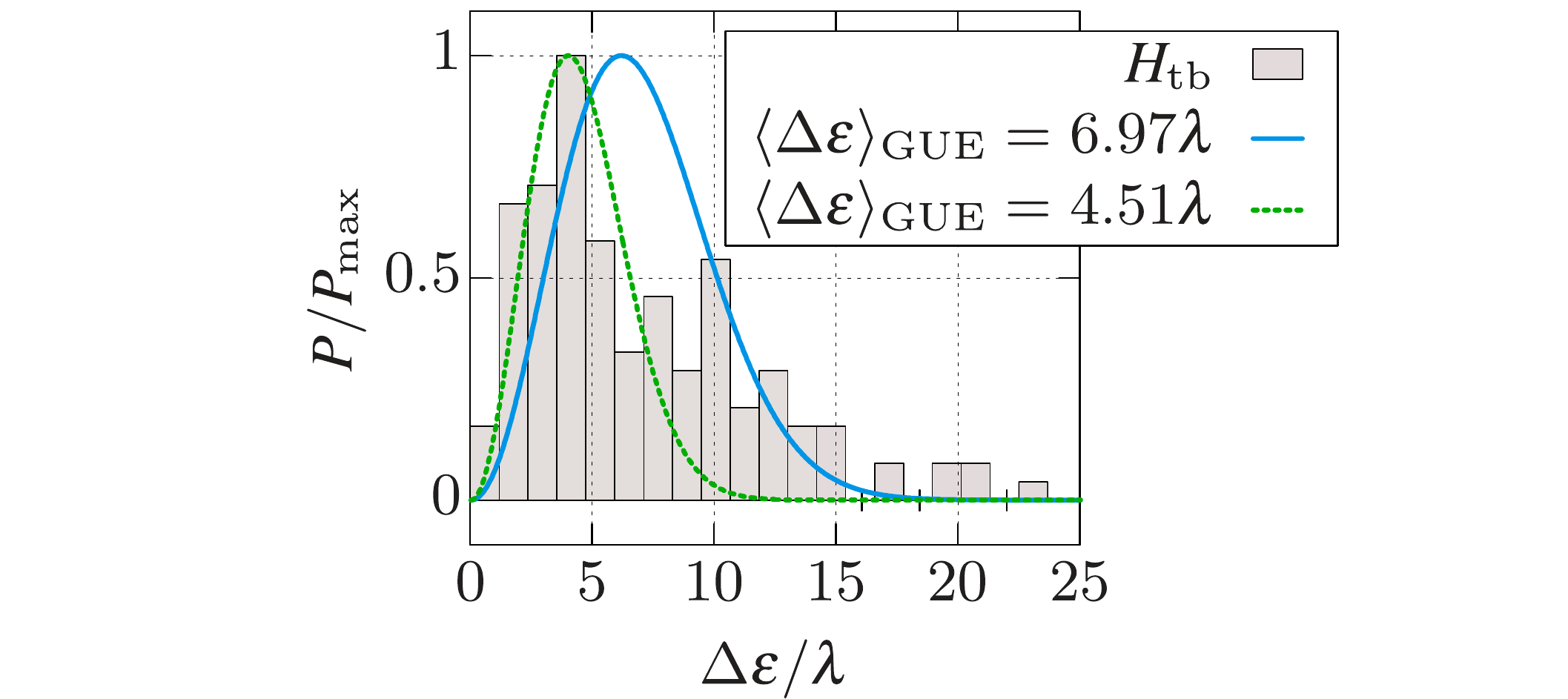}
\caption{Histogram of the splittings displayed in \Fig{fig_phaseshift_splitting}(e). We divide the distribution into 20 bins from $\Delta\epsilon = 0$ to $\Delta\epsilon = 25\lambda$ and normalize to the bin with the largest number of samples $P_{\text{max}}$. The blue and green curve show Gaussian unitary ensembles (GUEs) $P(\Delta\epsilon) \sim (\Delta\epsilon/\langle\Delta\epsilon\rangle)^2e^{-\frac{4}{\pi}\nbrack{\Delta\epsilon/\langle\Delta\epsilon\rangle}^2}$ with averages $\langle\Delta\epsilon\rangle_{\text{GUE}}$, normalized to their respective maximum values $P_{\text{max}}$ at $\Delta\epsilon = \sqrt{\pi}\langle\Delta\epsilon\rangle_{\text{GUE}}/2$. \label{fig_hist}}
\end{figure}

To diagonalize the Hamiltonian \eq{eq_hamiltonian_well_continuous}, we discretize it onto a rectangular lattice with discretization length $l$. The local basis states $\ket{\Psi^\sigma_{x,y}}$ are substituted by $\ket{\Psi^\sigma_{n_x,n_y}}$, signifying an electron with spin-z projection $\sigma = \uparrow\downarrow \equiv \pm1$ occupying the lattice site at $(x,y) = l(n_x,n_y)$ with $n_x,n_y\in\mathds{N}^0$. Derivatives of $\ket{\Psi^\sigma_{x,y}}$ with respect to $x,y$ become finite differences,
\begin{align}
 \partial_x\ket{\Psi^\sigma_{x,y}} &\rightarrow \frac{\ket{\Psi^\sigma_{n_x+1,n_y}} - \ket{\Psi^\sigma_{n_x-1,n_y}}}{2l}\notag\\
 \partial_y\ket{\Psi^\sigma_{x,y}} &\rightarrow \frac{\ket{\Psi^\sigma_{n_x,n_y+1}} - \ket{\Psi^\sigma_{n_x,n_y-1}}}{2l}\notag\\
 \partial^2_x\ket{\Psi^\sigma_{x,y}} &\rightarrow \frac{\ket{\Psi^\sigma_{n_x+1,n_y}} + \ket{\Psi^\sigma_{n_x-1,n_y}} - 2\ket{\Psi^\sigma_{n_x,n_y}}}{l^2}\notag\\
 \partial^2_y\ket{\Psi^\sigma_{x,y}} &\rightarrow \frac{\ket{\Psi^\sigma_{n_x,n_y+1}} + \ket{\Psi^\sigma_{n_x,n_y-1}} - 2\ket{\Psi^\sigma_{n_x,n_y}}}{l^2}.
\end{align}

To discretize the potential $V(x,y)$ in \eq{eq_hamiltonian_well_continuous} to $V_{n_x,n_y} = V(ln_x,ln_y)$, we bilinearly interpolate the $200\times 400$ pixel source image shown in \Fig{fig_phaseshift_splitting}(a) of the main text along the x and y direction.
Altogether, this maps \eq{eq_hamiltonian_well_continuous} to the tight-binding Hamiltonian
\begin{align}
 H_{\text{tb}} &= \sumsub{n_x,n_y}{\sigma=\pm}H^{\sigma,\vec{B}}_{n_x,n_y}
 \end{align}
 with 
 \begin{align}
 H^{\sigma,\vec{B}}_{n_x,n_y} &= \epsilon^{\sigma,B_z}_{n_x,n_y}\ket{\Psi^\sigma_{n_x,n_y}}\bra{\Psi^\sigma_{n_x,n_y}} + \frac{B_x}{2}\ket{\Psi^\sigma_{n_x,n_y}}\bra{\Psi^{-\sigma}_{n_x,n_y}}\notag\\
 &\phantom{=}-\sqlbrack{\tau^{B_z}_{n_y}\ket{\Psi^\sigma_{n_x,n_y}}\bra{\Psi^\sigma_{n_x + 1,n_y}}}\notag\\
 &\phantom{=-\sqlbrack{}}\sqrbrack{+ \tau^{B_z}_{n_x}\ket{\Psi^\sigma_{n_x,n_y}}\bra{\Psi^\sigma_{n_x,n_y + 1}} + \text{H.c.} }\notag\\
 &\phantom{=}- J\sqlbrack{\sigma \ket{\Psi^\sigma_{n_x,n_y}}\bra{\Psi^{-\sigma}_{n_x + 1,n_y}}}\notag\\
 &\phantom{=- J\sqlbrack{}}\sqrbrack{-i\ket{\Psi^\sigma_{n_x,n_y}}\bra{\Psi^{-\sigma}_{n_x,n_y+1}} + \text{H.c.}}.\label{eq_hamiltonian_well_tight_binding}
\end{align}

This includes
\begin{align}
 \epsilon^{\sigma,B_z}_{n_x,n_y} &= 4\tau + V_{n_x,n_y} + \frac{B_z\sigma}{2} + \tau\nbrack{\frac{m_e}{m^*}\frac{B_z}{2g_{\text{L}}\tau}}^2\nbrack{n_x^2 + n_y^2}\notag\\
 \tau^{B_z}_{n_x} &= \tau - i\frac{m_e}{m^*}\frac{B_z n_x}{2g_{\text{L}}} \lrsepa \tau^{B_z}_{n_y} = \tau + i\frac{m_e}{m^*}\frac{B_z n_y}{2g_{\text{L}}}.
\end{align}
The onsite-energy $\epsilon^{\sigma,B_z}_{n_x,n_y}$ consists of a kinetic contribution $4\tau$ given by the hopping amplitude $\tau = 1/(2m^*l^2)$, the potential energy $V_{n_x,n_y} = V(ln_x,ln_y)$, the spin-z Zeeman energy $B_z\sigma$, and the diamagnetic term $(A_x^2 + A_y^2)/(2m^*)$. The spin-conserving hopping amplitudes $\tau^{B_z}_{n_x}$
acquire a Peierls phase due to the perpendicular magnetic flux $\sim B_z$. The spin-orbit-coupling induced spin-flip hopping amplitude is given by $J = \alpha/(2l)$.

To obtain the approximate energies $\epsilon_j$ and eigenstates $\ket{\epsilon_j}$ of the dot Hamiltonian \eq{eq_hamiltonian_well_continuous}, we encode \eq{eq_hamiltonian_well_tight_binding} as a sparse matrix that we partially diagonalize with the Lanczos method described in \Refe{Simon1984}. More precisely, we only consider energy eigenstates of the Hamiltonian \eq{eq_hamiltonian_well_tight_binding} up to kinetic energies
 $0 < E_{\text{kin}} = \bra{\epsilon_j}H_{\text{tb}}(J,\vec{B},V_{n_x,n_y} = 0)\ket{\epsilon_j} \leq \tau/50$,
i.e., consistent with wavelengths $\sim (10/\sqrt{2}) l$, more than $5$ times larger than the discretization length. The distribution of nearest-neighbor splittings $\Delta\epsilon_j = \epsilon_j - \epsilon_{j-1}$ in this subspace is shown in \Fig{fig_phaseshift_splitting}(e) as a function of base level $\epsilon_{j-1}$, and histogrammed in \Fig{fig_hist}. The latter shows that for splittings up to $\Delta\epsilon \leq 5\lambda$, the distribution roughly follows a Wigner surmise obtained from a Gaussian unitary ensemble (GUE) with average $\langle\Delta\epsilon\rangle \approx 4.5\lambda$, c. f. green curve in \Fig{fig_hist}. However, a second peak around $\Delta\epsilon \approx 10\lambda$ shifts the average towards $\langle\Delta\epsilon\rangle \approx 6.97\lambda$. This clearly deviates from a Wigner surmise with equal average, see blue curve in \Fig{fig_hist}. Nevertheless, a clear energy level repulsion appears in the system.

\end{document}